\documentclass[runningheads]{llncs}
\usepackage[T1]{fontenc}
\usepackage{graphicx}
\usepackage{latexsym}
\usepackage{algorithm}
\usepackage[noend]{algorithmic}
\usepackage{amssymb}
\usepackage{amsmath}
\usepackage{booktabs}
\usepackage[inline]{enumitem}
\usepackage{graphicx}
\usepackage{hyperref}
\usepackage{environ}
\usepackage{marginnote}
\usepackage{multirow}
\usepackage{pgfplots}
\usepackage{pifont}
\usepackage{placeins}
\usepackage{rotate} 
\usepackage[draft]{todonotes}
\usepackage{tikz}
\usepackage{times} 
\usepackage{url}
\usepackage{varwidth}
\usepackage{verbatim} 
\usepackage{wrapfig}
\usepackage{xcolor,colortbl}
\usepackage{xspace}
\usepackage{subcaption}
\usepackage{tabularx}
\usetikzlibrary{intersections, backgrounds, calc}
\usepackage{cleveref}
\usepackage{color}

\begin{document}




\definecolor{snow}                {rgb}{1.00,0.98,0.98}
\definecolor{ghostwhite}          {rgb}{0.97,0.97,1.00}
\definecolor{whitesmoke}          {rgb}{0.96,0.96,0.96}
\definecolor{gainsboro}           {rgb}{0.86,0.86,0.86}
\definecolor{floralwhite}         {rgb}{1.00,0.98,0.94}
\definecolor{oldlace}             {rgb}{0.99,0.96,0.90}
\definecolor{linen}               {rgb}{0.98,0.94,0.90}
\definecolor{antiquewhite}        {rgb}{0.98,0.92,0.84}
\definecolor{papayawhip}          {rgb}{1.00,0.94,0.84}
\definecolor{blanchedalmond}      {rgb}{1.00,0.92,0.80}
\definecolor{bisque}              {rgb}{1.00,0.89,0.77}
\definecolor{peachpuff}           {rgb}{1.00,0.85,0.73}
\definecolor{navajowhite}         {rgb}{1.00,0.87,0.68}
\definecolor{moccasin}            {rgb}{1.00,0.89,0.71}
\definecolor{cornsilk}            {rgb}{1.00,0.97,0.86}
\definecolor{ivory}               {rgb}{1.00,1.00,0.94}
\definecolor{lemonchiffon}        {rgb}{1.00,0.98,0.80}
\definecolor{seashell}            {rgb}{1.00,0.96,0.93}
\definecolor{honeydew}            {rgb}{0.94,1.00,0.94}
\definecolor{mintcream}           {rgb}{0.96,1.00,0.98}
\definecolor{azure}               {rgb}{0.94,1.00,1.00}
\definecolor{aliceblue}           {rgb}{0.94,0.97,1.00}
\definecolor{lavender}            {rgb}{0.90,0.90,0.98}
\definecolor{lavenderblush}       {rgb}{1.00,0.94,0.96}
\definecolor{mistyrose}           {rgb}{1.00,0.89,0.88}
\definecolor{white}               {rgb}{1.00,1.00,1.00}
\definecolor{black}               {rgb}{0.00,0.00,0.00}
\definecolor{darkslategray}       {rgb}{0.18,0.31,0.31}
\definecolor{dimgray}             {rgb}{0.41,0.41,0.41}
\definecolor{slategray}           {rgb}{0.44,0.50,0.56}
\definecolor{lightslategray}      {rgb}{0.47,0.53,0.60}
\definecolor{gray}                {rgb}{0.75,0.75,0.75}
\definecolor{lightgrey}           {rgb}{0.83,0.83,0.83}
\definecolor{midnightblue}        {rgb}{0.10,0.10,0.44}
\definecolor{navy}                {rgb}{0.00,0.00,0.50}
\definecolor{cornflowerblue}      {rgb}{0.39,0.58,0.93}
\definecolor{darkslateblue}       {rgb}{0.28,0.24,0.55}
\definecolor{slateblue}           {rgb}{0.42,0.35,0.80}
\definecolor{mediumslateblue}     {rgb}{0.48,0.41,0.93}
\definecolor{lightslateblue}      {rgb}{0.52,0.44,1.00}
\definecolor{mediumblue}          {rgb}{0.00,0.00,0.80}
\definecolor{royalblue}           {rgb}{0.25,0.41,0.88}
\definecolor{blue}                {rgb}{0.00,0.00,1.00}
\definecolor{dodgerblue}          {rgb}{0.12,0.56,1.00}
\definecolor{deepskyblue}         {rgb}{0.00,0.75,1.00}
\definecolor{skyblue}             {rgb}{0.53,0.81,0.92}
\definecolor{lightskyblue}        {rgb}{0.53,0.81,0.98}
\definecolor{steelblue}           {rgb}{0.27,0.51,0.71}
\definecolor{lightsteelblue}      {rgb}{0.69,0.77,0.87}
\definecolor{lightblue}           {rgb}{0.68,0.85,0.90}
\definecolor{powderblue}          {rgb}{0.69,0.88,0.90}
\definecolor{paleturquoise}       {rgb}{0.69,0.93,0.93}
\definecolor{darkturquoise}       {rgb}{0.00,0.81,0.82}
\definecolor{mediumturquoise}     {rgb}{0.28,0.82,0.80}
\definecolor{turquoise}           {rgb}{0.25,0.88,0.82}
\definecolor{cyan}                {rgb}{0.00,1.00,1.00}
\definecolor{lightcyan}           {rgb}{0.88,1.00,1.00}
\definecolor{cadetblue}           {rgb}{0.37,0.62,0.63}
\definecolor{mediumaquamarine}    {rgb}{0.40,0.80,0.67}
\definecolor{aquamarine}          {rgb}{0.50,1.00,0.83}
\definecolor{darkgreen}           {rgb}{0.00,0.39,0.00}
\definecolor{darkolivegreen}      {rgb}{0.33,0.42,0.18}
\definecolor{darkseagreen}        {rgb}{0.56,0.74,0.56}
\definecolor{seagreen}            {rgb}{0.18,0.55,0.34}
\definecolor{mediumseagreen}      {rgb}{0.24,0.70,0.44}
\definecolor{lightseagreen}       {rgb}{0.13,0.70,0.67}
\definecolor{palegreen}           {rgb}{0.60,0.98,0.60}
\definecolor{springgreen}         {rgb}{0.00,1.00,0.50}
\definecolor{lawngreen}           {rgb}{0.49,0.99,0.00}
\definecolor{green}               {rgb}{0.00,1.00,0.00}
\definecolor{chartreuse}          {rgb}{0.50,1.00,0.00}
\definecolor{mediumspringgreen}   {rgb}{0.00,0.98,0.60}
\definecolor{greenyellow}         {rgb}{0.68,1.00,0.18}
\definecolor{limegreen}           {rgb}{0.20,0.80,0.20}
\definecolor{yellowgreen}         {rgb}{0.60,0.80,0.20}
\definecolor{forestgreen}         {rgb}{0.13,0.55,0.13}
\definecolor{olivedrab}           {rgb}{0.42,0.56,0.14}
\definecolor{darkkhaki}           {rgb}{0.74,0.72,0.42}
\definecolor{khaki}               {rgb}{0.94,0.90,0.55}
\definecolor{palegoldenrod}       {rgb}{0.93,0.91,0.67}
\definecolor{lightgoldenrodyellow} {rgb}{0.98,0.98,0.82}
\definecolor{lightyellow}         {rgb}{1.00,1.00,0.88}
\definecolor{yellow}              {rgb}{1.00,1.00,0.00}
\definecolor{gold}                {rgb}{1.00,0.84,0.00}
\definecolor{lightgoldenrod}      {rgb}{0.93,0.87,0.51}
\definecolor{goldenrod}           {rgb}{0.85,0.65,0.13}
\definecolor{darkgoldenrod}       {rgb}{0.72,0.53,0.04}
\definecolor{rosybrown}           {rgb}{0.74,0.56,0.56}
\definecolor{indianred}           {rgb}{0.80,0.36,0.36}
\definecolor{saddlebrown}         {rgb}{0.55,0.27,0.07}
\definecolor{sienna}              {rgb}{0.63,0.32,0.18}
\definecolor{peru}                {rgb}{0.80,0.52,0.25}
\definecolor{burlywood}           {rgb}{0.87,0.72,0.53}
\definecolor{beige}               {rgb}{0.96,0.96,0.86}
\definecolor{wheat}               {rgb}{0.96,0.87,0.70}
\definecolor{sandybrown}          {rgb}{0.96,0.64,0.38}
\definecolor{tan}                 {rgb}{0.82,0.71,0.55}
\definecolor{chocolate}           {rgb}{0.82,0.41,0.12}
\definecolor{firebrick}           {rgb}{0.70,0.13,0.13}
\definecolor{brown}               {rgb}{0.65,0.16,0.16}
\definecolor{darksalmon}          {rgb}{0.91,0.59,0.48}
\definecolor{salmon}              {rgb}{0.98,0.50,0.45}
\definecolor{lightsalmon}         {rgb}{1.00,0.63,0.48}
\definecolor{orange}              {rgb}{1.00,0.65,0.00}
\definecolor{darkorange}          {rgb}{1.00,0.55,0.00}
\definecolor{coral}               {rgb}{1.00,0.50,0.31}
\definecolor{lightcoral}          {rgb}{0.94,0.50,0.50}
\definecolor{tomato}              {rgb}{1.00,0.39,0.28}
\definecolor{orangered}           {rgb}{1.00,0.27,0.00}
\definecolor{red}                 {rgb}{1.00,0.00,0.00}
\definecolor{hotpink}             {rgb}{1.00,0.41,0.71}
\definecolor{deeppink}            {rgb}{1.00,0.08,0.58}
\definecolor{pink}                {rgb}{1.00,0.75,0.80}
\definecolor{lightpink}           {rgb}{1.00,0.71,0.76}
\definecolor{palevioletred}       {rgb}{0.86,0.44,0.58}
\definecolor{maroon}              {rgb}{0.69,0.19,0.38}
\definecolor{mediumvioletred}     {rgb}{0.78,0.08,0.52}
\definecolor{violetred}           {rgb}{0.82,0.13,0.56}
\definecolor{magenta}             {rgb}{1.00,0.00,1.00}
\definecolor{violet}              {rgb}{0.93,0.51,0.93}
\definecolor{plum}                {rgb}{0.87,0.63,0.87}
\definecolor{orchid}              {rgb}{0.85,0.44,0.84}
\definecolor{mediumorchid}        {rgb}{0.73,0.33,0.83}
\definecolor{darkorchid}          {rgb}{0.60,0.20,0.80}
\definecolor{darkviolet}          {rgb}{0.58,0.00,0.83}
\definecolor{blueviolet}          {rgb}{0.54,0.17,0.89}
\definecolor{purple}              {rgb}{0.63,0.13,0.94}
\definecolor{mediumpurple}        {rgb}{0.58,0.44,0.86}
\definecolor{thistle}             {rgb}{0.85,0.75,0.85}
\definecolor{snow2}               {rgb}{0.93,0.91,0.91}
\definecolor{snow3}               {rgb}{0.80,0.79,0.79}
\definecolor{snow4}               {rgb}{0.55,0.54,0.54}
\definecolor{seashell2}           {rgb}{0.93,0.90,0.87}
\definecolor{seashell3}           {rgb}{0.80,0.77,0.75}
\definecolor{seashell4}           {rgb}{0.55,0.53,0.51}
\definecolor{antiquewhite1}       {rgb}{1.00,0.94,0.86}
\definecolor{antiquewhite2}       {rgb}{0.93,0.87,0.80}
\definecolor{antiquewhite3}       {rgb}{0.80,0.75,0.69}
\definecolor{antiquewhite4}       {rgb}{0.55,0.51,0.47}
\definecolor{bisque2}             {rgb}{0.93,0.84,0.72}
\definecolor{bisque3}             {rgb}{0.80,0.72,0.62}
\definecolor{bisque4}             {rgb}{0.55,0.49,0.42}
\definecolor{peachpuff2}          {rgb}{0.93,0.80,0.68}
\definecolor{peachpuff3}          {rgb}{0.80,0.69,0.58}
\definecolor{peachpuff4}          {rgb}{0.55,0.47,0.40}
\definecolor{navajowhite2}        {rgb}{0.93,0.81,0.63}
\definecolor{navajowhite3}        {rgb}{0.80,0.70,0.55}
\definecolor{navajowhite4}        {rgb}{0.55,0.47,0.37}
\definecolor{lemonchiffon2}       {rgb}{0.93,0.91,0.75}
\definecolor{lemonchiffon3}       {rgb}{0.80,0.79,0.65}
\definecolor{lemonchiffon4}       {rgb}{0.55,0.54,0.44}
\definecolor{cornsilk2}           {rgb}{0.93,0.91,0.80}
\definecolor{cornsilk3}           {rgb}{0.80,0.78,0.69}
\definecolor{cornsilk4}           {rgb}{0.55,0.53,0.47}
\definecolor{ivory2}              {rgb}{0.93,0.93,0.88}
\definecolor{ivory3}              {rgb}{0.80,0.80,0.76}
\definecolor{ivory4}              {rgb}{0.55,0.55,0.51}
\definecolor{honeydew2}           {rgb}{0.88,0.93,0.88}
\definecolor{honeydew3}           {rgb}{0.76,0.80,0.76}
\definecolor{honeydew4}           {rgb}{0.51,0.55,0.51}
\definecolor{lavenderblush2}      {rgb}{0.93,0.88,0.90}
\definecolor{lavenderblush3}      {rgb}{0.80,0.76,0.77}
\definecolor{lavenderblush4}      {rgb}{0.55,0.51,0.53}
\definecolor{mistyrose2}          {rgb}{0.93,0.84,0.82}
\definecolor{mistyrose3}          {rgb}{0.80,0.72,0.71}
\definecolor{mistyrose4}          {rgb}{0.55,0.49,0.48}
\definecolor{azure2}              {rgb}{0.88,0.93,0.93}
\definecolor{azure3}              {rgb}{0.76,0.80,0.80}
\definecolor{azure4}              {rgb}{0.51,0.55,0.55}
\definecolor{slateblue1}          {rgb}{0.51,0.44,1.00}
\definecolor{slateblue2}          {rgb}{0.48,0.40,0.93}
\definecolor{slateblue3}          {rgb}{0.41,0.35,0.80}
\definecolor{slateblue4}          {rgb}{0.28,0.24,0.55}
\definecolor{royalblue1}          {rgb}{0.28,0.46,1.00}
\definecolor{royalblue2}          {rgb}{0.26,0.43,0.93}
\definecolor{royalblue3}          {rgb}{0.23,0.37,0.80}
\definecolor{royalblue4}          {rgb}{0.15,0.25,0.55}
\definecolor{blue2}               {rgb}{0.00,0.00,0.93}
\definecolor{blue4}               {rgb}{0.00,0.00,0.55}
\definecolor{dodgerblue2}         {rgb}{0.11,0.53,0.93}
\definecolor{dodgerblue3}         {rgb}{0.09,0.45,0.80}
\definecolor{dodgerblue4}         {rgb}{0.06,0.31,0.55}
\definecolor{steelblue1}          {rgb}{0.39,0.72,1.00}
\definecolor{steelblue2}          {rgb}{0.36,0.67,0.93}
\definecolor{steelblue3}          {rgb}{0.31,0.58,0.80}
\definecolor{steelblue4}          {rgb}{0.21,0.39,0.55}
\definecolor{deepskyblue2}        {rgb}{0.00,0.70,0.93}
\definecolor{deepskyblue3}        {rgb}{0.00,0.60,0.80}
\definecolor{deepskyblue4}        {rgb}{0.00,0.41,0.55}
\definecolor{skyblue1}            {rgb}{0.53,0.81,1.00}
\definecolor{skyblue2}            {rgb}{0.49,0.75,0.93}
\definecolor{skyblue3}            {rgb}{0.42,0.65,0.80}
\definecolor{skyblue4}            {rgb}{0.29,0.44,0.55}
\definecolor{lightskyblue1}       {rgb}{0.69,0.89,1.00}
\definecolor{lightskyblue2}       {rgb}{0.64,0.83,0.93}
\definecolor{lightskyblue3}       {rgb}{0.55,0.71,0.80}
\definecolor{lightskyblue4}       {rgb}{0.38,0.48,0.55}
\definecolor{slategray1}          {rgb}{0.78,0.89,1.00}
\definecolor{slategray2}          {rgb}{0.73,0.83,0.93}
\definecolor{slategray3}          {rgb}{0.62,0.71,0.80}
\definecolor{slategray4}          {rgb}{0.42,0.48,0.55}
\definecolor{lightsteelblue1}     {rgb}{0.79,0.88,1.00}
\definecolor{lightsteelblue2}     {rgb}{0.74,0.82,0.93}
\definecolor{lightsteelblue3}     {rgb}{0.64,0.71,0.80}
\definecolor{lightsteelblue4}     {rgb}{0.43,0.48,0.55}
\definecolor{lightblue1}          {rgb}{0.75,0.94,1.00}
\definecolor{lightblue2}          {rgb}{0.70,0.87,0.93}
\definecolor{lightblue3}          {rgb}{0.60,0.75,0.80}
\definecolor{lightblue4}          {rgb}{0.41,0.51,0.55}
\definecolor{lightcyan2}          {rgb}{0.82,0.93,0.93}
\definecolor{lightcyan3}          {rgb}{0.71,0.80,0.80}
\definecolor{lightcyan4}          {rgb}{0.48,0.55,0.55}
\definecolor{paleturquoise1}      {rgb}{0.73,1.00,1.00}
\definecolor{paleturquoise2}      {rgb}{0.68,0.93,0.93}
\definecolor{paleturquoise3}      {rgb}{0.59,0.80,0.80}
\definecolor{paleturquoise4}      {rgb}{0.40,0.55,0.55}
\definecolor{cadetblue1}          {rgb}{0.60,0.96,1.00}
\definecolor{cadetblue2}          {rgb}{0.56,0.90,0.93}
\definecolor{cadetblue3}          {rgb}{0.48,0.77,0.80}
\definecolor{cadetblue4}          {rgb}{0.33,0.53,0.55}
\definecolor{turquoise1}          {rgb}{0.00,0.96,1.00}
\definecolor{turquoise2}          {rgb}{0.00,0.90,0.93}
\definecolor{turquoise3}          {rgb}{0.00,0.77,0.80}
\definecolor{turquoise4}          {rgb}{0.00,0.53,0.55}
\definecolor{cyan2}               {rgb}{0.00,0.93,0.93}
\definecolor{cyan3}               {rgb}{0.00,0.80,0.80}
\definecolor{cyan4}               {rgb}{0.00,0.55,0.55}
\definecolor{darkslategray1}      {rgb}{0.59,1.00,1.00}
\definecolor{darkslategray2}      {rgb}{0.55,0.93,0.93}
\definecolor{darkslategray3}      {rgb}{0.47,0.80,0.80}
\definecolor{darkslategray4}      {rgb}{0.32,0.55,0.55}
\definecolor{aquamarine2}         {rgb}{0.46,0.93,0.78}
\definecolor{aquamarine4}         {rgb}{0.27,0.55,0.45}
\definecolor{darkseagreen1}       {rgb}{0.76,1.00,0.76}
\definecolor{darkseagreen2}       {rgb}{0.71,0.93,0.71}
\definecolor{darkseagreen3}       {rgb}{0.61,0.80,0.61}
\definecolor{darkseagreen4}       {rgb}{0.41,0.55,0.41}
\definecolor{seagreen1}           {rgb}{0.33,1.00,0.62}
\definecolor{seagreen2}           {rgb}{0.31,0.93,0.58}
\definecolor{seagreen3}           {rgb}{0.26,0.80,0.50}
\definecolor{palegreen1}          {rgb}{0.60,1.00,0.60}
\definecolor{palegreen2}          {rgb}{0.56,0.93,0.56}
\definecolor{palegreen3}          {rgb}{0.49,0.80,0.49}
\definecolor{palegreen4}          {rgb}{0.33,0.55,0.33}
\definecolor{springgreen2}        {rgb}{0.00,0.93,0.46}
\definecolor{springgreen3}        {rgb}{0.00,0.80,0.40}
\definecolor{springgreen4}        {rgb}{0.00,0.55,0.27}
\definecolor{green2}              {rgb}{0.00,0.93,0.00}
\definecolor{green3}              {rgb}{0.00,0.80,0.00}
\definecolor{green4}              {rgb}{0.00,0.55,0.00}
\definecolor{chartreuse2}         {rgb}{0.46,0.93,0.00}
\definecolor{chartreuse3}         {rgb}{0.40,0.80,0.00}
\definecolor{chartreuse4}         {rgb}{0.27,0.55,0.00}
\definecolor{olivedrab1}          {rgb}{0.75,1.00,0.24}
\definecolor{olivedrab2}          {rgb}{0.70,0.93,0.23}
\definecolor{olivedrab4}          {rgb}{0.41,0.55,0.13}
\definecolor{darkolivegreen1}     {rgb}{0.79,1.00,0.44}
\definecolor{darkolivegreen2}     {rgb}{0.74,0.93,0.41}
\definecolor{darkolivegreen3}     {rgb}{0.64,0.80,0.35}
\definecolor{darkolivegreen4}     {rgb}{0.43,0.55,0.24}
\definecolor{khaki1}              {rgb}{1.00,0.96,0.56}
\definecolor{khaki2}              {rgb}{0.93,0.90,0.52}
\definecolor{khaki3}              {rgb}{0.80,0.78,0.45}
\definecolor{khaki4}              {rgb}{0.55,0.53,0.31}
\definecolor{lightgoldenrod1}     {rgb}{1.00,0.93,0.55}
\definecolor{lightgoldenrod2}     {rgb}{0.93,0.86,0.51}
\definecolor{lightgoldenrod3}     {rgb}{0.80,0.75,0.44}
\definecolor{lightgoldenrod4}     {rgb}{0.55,0.51,0.30}
\definecolor{lightyellow2}        {rgb}{0.93,0.93,0.82}
\definecolor{lightyellow3}        {rgb}{0.80,0.80,0.71}
\definecolor{lightyellow4}        {rgb}{0.55,0.55,0.48}
\definecolor{yellow2}             {rgb}{0.93,0.93,0.00}
\definecolor{yellow3}             {rgb}{0.80,0.80,0.00}
\definecolor{yellow4}             {rgb}{0.55,0.55,0.00}
\definecolor{gold2}               {rgb}{0.93,0.79,0.00}
\definecolor{gold3}               {rgb}{0.80,0.68,0.00}
\definecolor{gold4}               {rgb}{0.55,0.46,0.00}
\definecolor{goldenrod1}          {rgb}{1.00,0.76,0.15}
\definecolor{goldenrod2}          {rgb}{0.93,0.71,0.13}
\definecolor{goldenrod3}          {rgb}{0.80,0.61,0.11}
\definecolor{goldenrod4}          {rgb}{0.55,0.41,0.08}
\definecolor{darkgoldenrod1}      {rgb}{1.00,0.73,0.06}
\definecolor{darkgoldenrod2}      {rgb}{0.93,0.68,0.05}
\definecolor{darkgoldenrod3}      {rgb}{0.80,0.58,0.05}
\definecolor{darkgoldenrod4}      {rgb}{0.55,0.40,0.03}
\definecolor{rosybrown1}          {rgb}{1.00,0.76,0.76}
\definecolor{rosybrown2}          {rgb}{0.93,0.71,0.71}
\definecolor{rosybrown3}          {rgb}{0.80,0.61,0.61}
\definecolor{rosybrown4}          {rgb}{0.55,0.41,0.41}
\definecolor{indianred1}          {rgb}{1.00,0.42,0.42}
\definecolor{indianred2}          {rgb}{0.93,0.39,0.39}
\definecolor{indianred3}          {rgb}{0.80,0.33,0.33}
\definecolor{indianred4}          {rgb}{0.55,0.23,0.23}
\definecolor{sienna1}             {rgb}{1.00,0.51,0.28}
\definecolor{sienna2}             {rgb}{0.93,0.47,0.26}
\definecolor{sienna3}             {rgb}{0.80,0.41,0.22}
\definecolor{sienna4}             {rgb}{0.55,0.28,0.15}
\definecolor{burlywood1}          {rgb}{1.00,0.83,0.61}
\definecolor{burlywood2}          {rgb}{0.93,0.77,0.57}
\definecolor{burlywood3}          {rgb}{0.80,0.67,0.49}
\definecolor{burlywood4}          {rgb}{0.55,0.45,0.33}
\definecolor{wheat1}              {rgb}{1.00,0.91,0.73}
\definecolor{wheat2}              {rgb}{0.93,0.85,0.68}
\definecolor{wheat3}              {rgb}{0.80,0.73,0.59}
\definecolor{wheat4}              {rgb}{0.55,0.49,0.40}
\definecolor{tan1}                {rgb}{1.00,0.65,0.31}
\definecolor{tan2}                {rgb}{0.93,0.60,0.29}
\definecolor{tan4}                {rgb}{0.55,0.35,0.17}
\definecolor{chocolate1}          {rgb}{1.00,0.50,0.14}
\definecolor{chocolate2}          {rgb}{0.93,0.46,0.13}
\definecolor{chocolate3}          {rgb}{0.80,0.40,0.11}
\definecolor{firebrick1}          {rgb}{1.00,0.19,0.19}
\definecolor{firebrick2}          {rgb}{0.93,0.17,0.17}
\definecolor{firebrick3}          {rgb}{0.80,0.15,0.15}
\definecolor{firebrick4}          {rgb}{0.55,0.10,0.10}
\definecolor{brown1}              {rgb}{1.00,0.25,0.25}
\definecolor{brown2}              {rgb}{0.93,0.23,0.23}
\definecolor{brown3}              {rgb}{0.80,0.20,0.20}
\definecolor{brown4}              {rgb}{0.55,0.14,0.14}
\definecolor{salmon1}             {rgb}{1.00,0.55,0.41}
\definecolor{salmon2}             {rgb}{0.93,0.51,0.38}
\definecolor{salmon3}             {rgb}{0.80,0.44,0.33}
\definecolor{salmon4}             {rgb}{0.55,0.30,0.22}
\definecolor{lightsalmon2}        {rgb}{0.93,0.58,0.45}
\definecolor{lightsalmon3}        {rgb}{0.80,0.51,0.38}
\definecolor{lightsalmon4}        {rgb}{0.55,0.34,0.26}
\definecolor{orange2}             {rgb}{0.93,0.60,0.00}
\definecolor{orange3}             {rgb}{0.80,0.52,0.00}
\definecolor{orange4}             {rgb}{0.55,0.35,0.00}
\definecolor{darkorange1}         {rgb}{1.00,0.50,0.00}
\definecolor{darkorange2}         {rgb}{0.93,0.46,0.00}
\definecolor{darkorange3}         {rgb}{0.80,0.40,0.00}
\definecolor{darkorange4}         {rgb}{0.55,0.27,0.00}
\definecolor{coral1}              {rgb}{1.00,0.45,0.34}
\definecolor{coral2}              {rgb}{0.93,0.42,0.31}
\definecolor{coral3}              {rgb}{0.80,0.36,0.27}
\definecolor{coral4}              {rgb}{0.55,0.24,0.18}
\definecolor{tomato2}             {rgb}{0.93,0.36,0.26}
\definecolor{tomato3}             {rgb}{0.80,0.31,0.22}
\definecolor{tomato4}             {rgb}{0.55,0.21,0.15}
\definecolor{orangered2}          {rgb}{0.93,0.25,0.00}
\definecolor{orangered3}          {rgb}{0.80,0.22,0.00}
\definecolor{orangered4}          {rgb}{0.55,0.15,0.00}
\definecolor{red2}                {rgb}{0.93,0.00,0.00}
\definecolor{red3}                {rgb}{0.80,0.00,0.00}
\definecolor{red4}                {rgb}{0.55,0.00,0.00}
\definecolor{deeppink2}           {rgb}{0.93,0.07,0.54}
\definecolor{deeppink3}           {rgb}{0.80,0.06,0.46}
\definecolor{deeppink4}           {rgb}{0.55,0.04,0.31}
\definecolor{hotpink1}            {rgb}{1.00,0.43,0.71}
\definecolor{hotpink2}            {rgb}{0.93,0.42,0.65}
\definecolor{hotpink3}            {rgb}{0.80,0.38,0.56}
\definecolor{hotpink4}            {rgb}{0.55,0.23,0.38}
\definecolor{pink1}               {rgb}{1.00,0.71,0.77}
\definecolor{pink2}               {rgb}{0.93,0.66,0.72}
\definecolor{pink3}               {rgb}{0.80,0.57,0.62}
\definecolor{pink4}               {rgb}{0.55,0.39,0.42}
\definecolor{lightpink1}          {rgb}{1.00,0.68,0.73}
\definecolor{lightpink2}          {rgb}{0.93,0.64,0.68}
\definecolor{lightpink3}          {rgb}{0.80,0.55,0.58}
\definecolor{lightpink4}          {rgb}{0.55,0.37,0.40}
\definecolor{palevioletred1}      {rgb}{1.00,0.51,0.67}
\definecolor{palevioletred2}      {rgb}{0.93,0.47,0.62}
\definecolor{palevioletred3}      {rgb}{0.80,0.41,0.54}
\definecolor{palevioletred4}      {rgb}{0.55,0.28,0.36}
\definecolor{maroon1}             {rgb}{1.00,0.20,0.70}
\definecolor{maroon2}             {rgb}{0.93,0.19,0.65}
\definecolor{maroon3}             {rgb}{0.80,0.16,0.56}
\definecolor{maroon4}             {rgb}{0.55,0.11,0.38}
\definecolor{violetred1}          {rgb}{1.00,0.24,0.59}
\definecolor{violetred2}          {rgb}{0.93,0.23,0.55}
\definecolor{violetred3}          {rgb}{0.80,0.20,0.47}
\definecolor{violetred4}          {rgb}{0.55,0.13,0.32}
\definecolor{magenta2}            {rgb}{0.93,0.00,0.93}
\definecolor{magenta3}            {rgb}{0.80,0.00,0.80}
\definecolor{magenta4}            {rgb}{0.55,0.00,0.55}
\definecolor{orchid1}             {rgb}{1.00,0.51,0.98}
\definecolor{orchid2}             {rgb}{0.93,0.48,0.91}
\definecolor{orchid3}             {rgb}{0.80,0.41,0.79}
\definecolor{orchid4}             {rgb}{0.55,0.28,0.54}
\definecolor{plum1}               {rgb}{1.00,0.73,1.00}
\definecolor{plum2}               {rgb}{0.93,0.68,0.93}
\definecolor{plum3}               {rgb}{0.80,0.59,0.80}
\definecolor{plum4}               {rgb}{0.55,0.40,0.55}
\definecolor{mediumorchid1}       {rgb}{0.88,0.40,1.00}
\definecolor{mediumorchid2}       {rgb}{0.82,0.37,0.93}
\definecolor{mediumorchid3}       {rgb}{0.71,0.32,0.80}
\definecolor{mediumorchid4}       {rgb}{0.48,0.22,0.55}
\definecolor{darkorchid1}         {rgb}{0.75,0.24,1.00}
\definecolor{darkorchid2}         {rgb}{0.70,0.23,0.93}
\definecolor{darkorchid3}         {rgb}{0.60,0.20,0.80}
\definecolor{darkorchid4}         {rgb}{0.41,0.13,0.55}
\definecolor{purple1}             {rgb}{0.61,0.19,1.00}
\definecolor{purple2}             {rgb}{0.57,0.17,0.93}
\definecolor{purple3}             {rgb}{0.49,0.15,0.80}
\definecolor{purple4}             {rgb}{0.33,0.10,0.55}
\definecolor{mediumpurple1}       {rgb}{0.67,0.51,1.00}
\definecolor{mediumpurple2}       {rgb}{0.62,0.47,0.93}
\definecolor{mediumpurple3}       {rgb}{0.54,0.41,0.80}
\definecolor{mediumpurple4}       {rgb}{0.36,0.28,0.55}
\definecolor{thistle1}            {rgb}{1.00,0.88,1.00}
\definecolor{thistle2}            {rgb}{0.93,0.82,0.93}
\definecolor{thistle3}            {rgb}{0.80,0.71,0.80}
\definecolor{thistle4}            {rgb}{0.55,0.48,0.55}
\definecolor{gray1}               {rgb}{0.01,0.01,0.01}
\definecolor{gray2}               {rgb}{0.02,0.02,0.02}
\definecolor{gray3}               {rgb}{0.03,0.03,0.03}
\definecolor{gray4}               {rgb}{0.04,0.04,0.04}
\definecolor{gray5}               {rgb}{0.05,0.05,0.05}
\definecolor{gray6}               {rgb}{0.06,0.06,0.06}
\definecolor{gray7}               {rgb}{0.07,0.07,0.07}
\definecolor{gray8}               {rgb}{0.08,0.08,0.08}
\definecolor{gray9}               {rgb}{0.09,0.09,0.09}
\definecolor{gray10}              {rgb}{0.10,0.10,0.10}
\definecolor{gray11}              {rgb}{0.11,0.11,0.11}
\definecolor{gray12}              {rgb}{0.12,0.12,0.12}
\definecolor{gray13}              {rgb}{0.13,0.13,0.13}
\definecolor{gray14}              {rgb}{0.14,0.14,0.14}
\definecolor{gray15}              {rgb}{0.15,0.15,0.15}
\definecolor{gray16}              {rgb}{0.16,0.16,0.16}
\definecolor{gray17}              {rgb}{0.17,0.17,0.17}
\definecolor{gray18}              {rgb}{0.18,0.18,0.18}
\definecolor{gray19}              {rgb}{0.19,0.19,0.19}
\definecolor{gray20}              {rgb}{0.20,0.20,0.20}
\definecolor{gray21}              {rgb}{0.21,0.21,0.21}
\definecolor{gray22}              {rgb}{0.22,0.22,0.22}
\definecolor{gray23}              {rgb}{0.23,0.23,0.23}
\definecolor{gray24}              {rgb}{0.24,0.24,0.24}
\definecolor{gray25}              {rgb}{0.25,0.25,0.25}
\definecolor{gray26}              {rgb}{0.26,0.26,0.26}
\definecolor{gray27}              {rgb}{0.27,0.27,0.27}
\definecolor{gray28}              {rgb}{0.28,0.28,0.28}
\definecolor{gray29}              {rgb}{0.29,0.29,0.29}
\definecolor{gray30}              {rgb}{0.30,0.30,0.30}
\definecolor{gray31}              {rgb}{0.31,0.31,0.31}
\definecolor{gray32}              {rgb}{0.32,0.32,0.32}
\definecolor{gray33}              {rgb}{0.33,0.33,0.33}
\definecolor{gray34}              {rgb}{0.34,0.34,0.34}
\definecolor{gray35}              {rgb}{0.35,0.35,0.35}
\definecolor{gray36}              {rgb}{0.36,0.36,0.36}
\definecolor{gray37}              {rgb}{0.37,0.37,0.37}
\definecolor{gray38}              {rgb}{0.38,0.38,0.38}
\definecolor{gray39}              {rgb}{0.39,0.39,0.39}
\definecolor{gray40}              {rgb}{0.40,0.40,0.40}
\definecolor{gray42}              {rgb}{0.42,0.42,0.42}
\definecolor{gray43}              {rgb}{0.43,0.43,0.43}
\definecolor{gray44}              {rgb}{0.44,0.44,0.44}
\definecolor{gray45}              {rgb}{0.45,0.45,0.45}
\definecolor{gray46}              {rgb}{0.46,0.46,0.46}
\definecolor{gray47}              {rgb}{0.47,0.47,0.47}
\definecolor{gray48}              {rgb}{0.48,0.48,0.48}
\definecolor{gray49}              {rgb}{0.49,0.49,0.49}
\definecolor{gray50}              {rgb}{0.50,0.50,0.50}
\definecolor{gray51}              {rgb}{0.51,0.51,0.51}
\definecolor{gray52}              {rgb}{0.52,0.52,0.52}
\definecolor{gray53}              {rgb}{0.53,0.53,0.53}
\definecolor{gray54}              {rgb}{0.54,0.54,0.54}
\definecolor{gray55}              {rgb}{0.55,0.55,0.55}
\definecolor{gray56}              {rgb}{0.56,0.56,0.56}
\definecolor{gray57}              {rgb}{0.57,0.57,0.57}
\definecolor{gray58}              {rgb}{0.58,0.58,0.58}
\definecolor{gray59}              {rgb}{0.59,0.59,0.59}
\definecolor{gray60}              {rgb}{0.60,0.60,0.60}
\definecolor{gray61}              {rgb}{0.61,0.61,0.61}
\definecolor{gray62}              {rgb}{0.62,0.62,0.62}
\definecolor{gray63}              {rgb}{0.63,0.63,0.63}
\definecolor{gray64}              {rgb}{0.64,0.64,0.64}
\definecolor{gray65}              {rgb}{0.65,0.65,0.65}
\definecolor{gray66}              {rgb}{0.66,0.66,0.66}
\definecolor{gray67}              {rgb}{0.67,0.67,0.67}
\definecolor{gray68}              {rgb}{0.68,0.68,0.68}
\definecolor{gray69}              {rgb}{0.69,0.69,0.69}
\definecolor{gray70}              {rgb}{0.70,0.70,0.70}
\definecolor{gray71}              {rgb}{0.71,0.71,0.71}
\definecolor{gray72}              {rgb}{0.72,0.72,0.72}
\definecolor{gray73}              {rgb}{0.73,0.73,0.73}
\definecolor{gray74}              {rgb}{0.74,0.74,0.74}
\definecolor{gray75}              {rgb}{0.75,0.75,0.75}
\definecolor{gray76}              {rgb}{0.76,0.76,0.76}
\definecolor{gray77}              {rgb}{0.77,0.77,0.77}
\definecolor{gray78}              {rgb}{0.78,0.78,0.78}
\definecolor{gray79}              {rgb}{0.79,0.79,0.79}
\definecolor{gray80}              {rgb}{0.80,0.80,0.80}
\definecolor{gray81}              {rgb}{0.81,0.81,0.81}
\definecolor{gray82}              {rgb}{0.82,0.82,0.82}
\definecolor{gray83}              {rgb}{0.83,0.83,0.83}
\definecolor{gray84}              {rgb}{0.84,0.84,0.84}
\definecolor{gray85}              {rgb}{0.85,0.85,0.85}
\definecolor{gray86}              {rgb}{0.86,0.86,0.86}
\definecolor{gray87}              {rgb}{0.87,0.87,0.87}
\definecolor{gray88}              {rgb}{0.88,0.88,0.88}
\definecolor{gray89}              {rgb}{0.89,0.89,0.89}
\definecolor{gray90}              {rgb}{0.90,0.90,0.90}
\definecolor{gray91}              {rgb}{0.91,0.91,0.91}
\definecolor{gray92}              {rgb}{0.92,0.92,0.92}
\definecolor{gray93}              {rgb}{0.93,0.93,0.93}
\definecolor{gray94}              {rgb}{0.94,0.94,0.94}
\definecolor{gray95}              {rgb}{0.95,0.95,0.95}
\definecolor{gray97}              {rgb}{0.97,0.97,0.97}
\definecolor{gray98}              {rgb}{0.98,0.98,0.98}
\definecolor{gray99}              {rgb}{0.99,0.99,0.99}
\definecolor{darkgrey}            {rgb}{0.66,0.66,0.66}

\renewcommand{\algorithmiccomment}[1]{\ \ \ \ \ // #1}


\newcommand{\RSHIGHLIGHT}[1]{\textcolor{blue}{{{#1}}}}

\newcommand{\new}[1]{{\blue #1}\/}
\newcommand{\ex}[1]{{\green #1}\/}

\newcommand{\resp}[1]{[resp.\ #1]}
\newcommand{\done}{\textcolor{darkgreen}{\checkmark}}

\newcommand{\TODO}[1]{{}}
\newcommand{\ACTODO}[1]{{\bf \textcolor{red}{{\fbox{AC TODO:} #1}}}}
\newcommand{\RS}[1]{\textcolor{blue}{#1}}
\newcommand{\ignore}[1]{}
\newcommand{\RSCHANGE}[1]{\textcolor{blue}{#1}}
\renewcommand{\RS}[1]{\textcolor{darkgreen}{#1}}
\newcommand{\PT}[1]{\textcolor{darkviolet}{#1}}
\newcommand{\RSTODO}[1]{{\bf \textcolor{darkgreen}{{\fbox{RS TODO:} #1}}}}
\renewcommand{\RSTODO}[1]{}
\newcommand{\nota}[1] {\noindent \fbox{{\bf NOTA:}} #1 }
\newcommand{\marg}[1]{\marginpar{\ \\{\em #1\/}}}
\renewcommand{\marg}[1]{\marginpar{\ \\\textcolor{blue}{\ {\sf #1\/}}}}

\newenvironment{rschange}{\color{darkviolet}}{\normalcolor}
\newenvironment{rs}{\color{blue}}{\normalcolor}
\newenvironment{pt}{\color{darkviolet}}{\normalcolor}

\newcommand{\RSNOTE}[1]{\marginpar{\textcolor{darkgreen}{\textbf{RS: }
            {\footnotesize #1}}}}
\renewcommand{\RSNOTE}[1]{\noindent\textcolor{darkgreen}{{\bf RS: #1}}}

\newcommand{\ignoreinshort}[1]{}
\newcommand{\ignoreinlong}[1]{{#1}}


%
%
\providecommand{\longversion}{true}
\ifthenelse{\equal{\longversion}{true}}
{%
    \renewcommand{\ignoreinshort}[1]{\textcolor{blue}{#1}}
    \newcommand{\ignoreinshortnc}[1]{{#1}}
    \renewcommand{\ignoreinlong}[1]{}
    \newcommand{\ignoreinlongnc}[1]{}
    \newenvironment{ignoreinshortenv}{\color{blue}}{\normalcolor}
    \NewEnviron{ignoreinlongenv}{}
}%
{%
    \renewcommand{\ignoreinshort}[1]{}
    \newcommand{\ignoreinshortnc}[1]{}
    \renewcommand{\ignoreinlong}[1]{\textcolor{blue}{#1}}
    \newcommand{\ignoreinlongnc}[1]{{#1}}
    \NewEnviron{ignoreinshortenv}{}
    \newenvironment{ignoreinlongenv}{\color{blue}}{\normalcolor}
}

\def\makenewenumerate#1#2{%
    \newcounter{cnt#1}
    \newenvironment{#1}%
    {\begin{list}{\makebox[0pt][r]{#2}}%
            {\setlength{\itemsep}{0pt}%
                \setlength{\parsep}{.2em}%
                \setlength{\leftmargin}{1.5em}%
                \setlength{\labelwidth}{.4em}%
                \usecounter{cnt#1}}}
            {\end{list}}}

\makenewenumerate{myenumerate}{\arabic{cntmyenumerate}.}

\makenewenumerate{renumerate}{\rm(\roman{cntrenumerate})}
\makenewenumerate{renumerateprime}{\rm(\roman{cntrenumerateprime}$'$)}
\makenewenumerate{renumeratesecond}{\rm(\roman{cntrenumeratesecond}$''$)}

\makenewenumerate{aenumerate}{({\it\alph{cntaenumerate}})}
\makenewenumerate{aenumerateprime}{({\it\alph{cntaenumerateprime}$'$})}
\makenewenumerate{aenumeratesecond}{({\it\alph{cntaenumeratesecond}$''$})}



\newcommand{\fakesubsubsection}[1]{\smallskip\noindent {\bf #1.}}


\newcommand{\sref}[1]{\S{}\ref{#1}}
\newcommand{\noi}{\noindent}


\newcommand{\pair}[2]{\ensuremath{\langle{#1},{#2}\rangle}\xspace}
\newcommand{\triple}[3]{\ensuremath{\langle{#1},{#2},{#3}\rangle}\xspace}
\newcommand{\tuple}[1]{\ensuremath{\langle{#1}\rangle}\xspace}
\newcommand{\set}[1]{\ensuremath{\{{#1}\}}\xspace}
\newcommand{\imp}{\ensuremath{\rightarrow}\xspace}
\newcommand{\limp}{\ensuremath{\leftarrow}\xspace}
\renewcommand{\iff}{\ensuremath{\leftrightarrow}\xspace}
\newcommand{\defas}{\ensuremath{\stackrel{\text{\scalebox{.7}{def}}}{=}}\xspace}
\newcommand{\thus}{\ensuremath{\Longrightarrow}\xspace}

\newcommand{\pos}{\phantom{\neg}}


\newcommand\cala{\ensuremath{\mathcal{A}}\xspace}
\newcommand\calb{\ensuremath{\mathcal{B}}\xspace}
\newcommand\calc{\ensuremath{\mathcal{C}}\xspace}
\newcommand\cald{\ensuremath{\mathcal{D}}\xspace}
\newcommand\cale{\ensuremath{\mathcal{E}}\xspace}
\newcommand\calf{\ensuremath{\mathcal{F}}\xspace}
\newcommand\calg{\ensuremath{\mathcal{G}}\xspace}
\newcommand\calh{\ensuremath{\mathcal{H}}\xspace}
\newcommand\cali{\ensuremath{\mathcal{I}}\xspace}
\newcommand\calj{\ensuremath{\mathcal{J}}\xspace}
\newcommand\calk{\ensuremath{\mathcal{K}}\xspace}
\newcommand\call{\ensuremath{\mathcal{L}}\xspace}
\newcommand\calm{\ensuremath{\mathcal{M}}\xspace}
\newcommand\caln{\ensuremath{\mathcal{N}}\xspace}
\newcommand\calo{\ensuremath{\mathcal{O}}\xspace}
\newcommand\calp{\ensuremath{\mathcal{P}}\xspace}
\newcommand\calq{\ensuremath{\mathcal{Q}}\xspace}
\newcommand\calr{\ensuremath{\mathcal{R}}\xspace}
\newcommand\cals{\ensuremath{\mathcal{S}}\xspace}
\newcommand\calt{\ensuremath{\mathcal{T}}\xspace}
\newcommand\calu{\ensuremath{\mathcal{U}}\xspace}
\newcommand\calv{\ensuremath{\mathcal{V}}\xspace}
\newcommand\calw{\ensuremath{\mathcal{W}}\xspace}
\newcommand\calx{\ensuremath{\mathcal{X}}\xspace}
\newcommand\caly{\ensuremath{\mathcal{Y}}\xspace}
\newcommand\calz{\ensuremath{\mathcal{Z}}\xspace}

\newcommand{\rone}{\marginnote{\ \textcolor{red}{\ {\sf R1\/}}}}
\newcommand{\rtwo}{\marginnote{\ \textcolor{red}{\ {\sf R2\/}}}}
\newcommand{\rthree}{\marginnote{\ \textcolor{red}{\ {\sf R3\/}}}}
\newcommand{\ronetwo}{\marginnote{\ \textcolor{red}{\ {\sf R1\/}, {\sf R2\/}}}}
\newcommand{\ronethree}{\marginnote{\ \textcolor{red}{\ {\sf R1\/}, {\sf R3\/}}}}

\newcommand{\ronefour}{\marginnote{\ \textcolor{red}{\ {\sf R1\/}, {\sf R4\/}}}}
\newcommand{\rtwothree}{\marginnote{\ \textcolor{red}{\ {\sf R2\/}, {\sf R3\/}}}}
\newcommand{\rtwofour}{\marginnote{\ \textcolor{red}{\ {\sf R2\/}, {\sf R4\/}}}}

\newcommand{\ronetwothree}{\marginnote{\ \textcolor{red}{\ {\sf
R1\/}, {\sf R2\/} {\sf R3\/}}}}
\newcommand{\ronethreefour}{\marginnote{\ \textcolor{red}{\ {\sf
R1\/}, {\sf R3\/} {\sf R4\/}}}}
\newcommand{\rtwothreefour}{\marginnote{\ \textcolor{red}{\ {\sf
R2\/}, {\sf R3\/} {\sf R4\/}}}}

\newcommand{\ronetwothreefour}{\marginnote{\ \textcolor{red}{\ {\sf
R1\/}, {\sf R2\/}, {\sf R3\/}  {\sf R4\/}}}}

\renewcommand{\RSCHANGE}[1]{\textcolor{darkviolet}{{#1}}}
\newcommand{\RSCHANGEONE}[1]{\rone{}~\textcolor{blue}{{#1}}}
\newcommand{\RSCHANGETWO}[1]{\rtwo{}~\textcolor{blue}{{#1}}}
\newcommand{\RSCHANGETHREE}[1]{\rthree{}~\textcolor{blue}{{#1}}}
\newcommand{\RSCHANGEONETHREE}[1]{\ronethree{}~\textcolor{blue}{{#1}}}
\newcommand{\RSCHANGEONETWO}[1]{\ronetwo{}~\textcolor{blue}{{#1}}}
\newcommand{\RSCHANGETWOTHREE}[1]{\rtwothree{}~\textcolor{blue}{{#1}}}

\newcommand{\RSCHANGEFOUR}[1]{\rfour{}~\textcolor{blue}{{#1}}}
\newcommand{\RSCHANGEONEFOUR}[1]{\ronefour{}~\textcolor{blue}{{#1}}}
\newcommand{\RSCHANGETWOFOUR}[1]{\rtwofour{}~\textcolor{blue}{{#1}}}
\newcommand{\RSCHANGETHREEFOUR}[1]{\rtwothreefour{}~\textcolor{blue}{{#1}}}
\newcommand{\RSCHANGEONETHREEFOUR}[1]{\ronethreefour{}~\textcolor{blue}{{#1}}}


\newcommand{\en}[2]{{#1}e{#2}}
\newcommand{\den}[4]{\en{#1}{#2}/\en{#3}{#4}}

\newcommand{\mc}[1]{\ensuremath{\mathcal{#1}}}
\newcommand{\msf}[1]{\ensuremath{\mathsf{#1}}}

\newcommand{\snomed}{\textsc{Snomed-CT}\xspace}
\newcommand{\geneonto}{\textsc{GeneOntology}\xspace}
\newcommand{\nci}{\textsc{NCI}\xspace}
\newcommand{\galen}{\textsc{Galen}\xspace}

\newcommand{\shsnomed}{\textsc{Snomed09}\xspace}
\newcommand{\shnci}{\textsc{NCI}\xspace}
\newcommand{\shgalen}{\textsc{notGalen}\xspace}
\newcommand{\shfullgalen}{{\textsc{fullGalen}}\xspace}
\newcommand{\shgeneontology}{{\textsc{GeneOnt.}}\xspace}

\newcommand{\conj}{\sqcap}
\newcommand{\disj}{\sqcup}
\newcommand{\exr}[2]{{\exists}#1.#2}

\newcommand{\subs}{\sqsubseteq}
\newcommand{\sups}{\sqsupseteq}
\newcommand{\norm}{\triangleq}
\newcommand{\comp}{\circ}

\newcommand{\topt}{\ensuremath{\top}\xspace}
\newcommand{\bott}{\ensuremath{\bot}\xspace}
\newcommand{\conjt}[2]{\ensuremath{#1 \sqcap #2}\xspace}
\newcommand{\disjt}[2]{\ensuremath{#1 \sqcup #2}\xspace}
\newcommand{\exrt}[2]{\ensuremath{{\exists}#1.#2}\xspace}
\newcommand{\equivt}[2]{\ensuremath{#1 \equiv #2}\xspace}
\newcommand{\subst}[2]{\ensuremath{#1 \sqsubseteq #2}\xspace}
\newcommand{\supst}[2]{\ensuremath{#1 \sqsupseteq #2}\xspace}

\newcommand{\pr}[1]{\ensuremath{p_{#1}}\xspace}
\newcommand{\prc}[1]{\ensuremath{p_{[#1]}}\xspace}
\newcommand{\sel}[1]{\ensuremath{s_{#1}}\xspace}
\newcommand{\selc}[1]{\ensuremath{s_{[#1]}}\xspace}

\newcommand{\goal}{\ensuremath{C_i\subs_\T D_i}\xspace}
\newcommand{\goalsure}{\ensuremath{C_i~\subs_\T~D_i}\xspace}
\newcommand{\tgoal}[1]{\ensuremath{C_i\subs_{#1} D_i}\xspace}
\newcommand{\igoal}[1]{\ensuremath{{C_{#1} \subs_\T D_{#1}}}\xspace}
\newcommand{\sgoal}{\ensuremath{\selc{\goal}}\xspace}
\newcommand{\isgoal}[1]{\ensuremath{\selc{\igoal{#1}}}\xspace}
\newcommand{\nsgoal}{\ensuremath{\neg\selc{C_i\subs_\T D_i}}\xspace}

\newcommand{\assumptiongoal}{\ensuremath{\{\neg\prc{D_i},\prc{C_i}\}}\xspace}
\newcommand{\assumptionsgoal}{\ensuremath{\{\neg\sgoal\}}\xspace}

\newcommand{\aenc}[1]{\ensuremath{\elp2sat(#1)}\xspace}
\newcommand{\wffsubs}{\ensuremath{\phi_{\T{}}}\xspace}
\newcommand{\wffsubsnotrans}{\ensuremath{\phi_{\T{}}^{notrans}}\xspace}
\newcommand{\wffmina}{\ensuremath{\wffsubs^{one}}\xspace}
\newcommand{\wffminarule}{\ensuremath{\phi_{\T(po)}^{one}}\xspace}
\newcommand{\wffminaT}[1]{\ensuremath{\phi_{#1}^{one}}\xspace}
\newcommand{\wffminaTrule}[1]{\ensuremath{\phi_{#1}^{one}}\xspace}
\newcommand{\wffallmina}{\ensuremath{\wffsubs^{all}}\xspace}
\newcommand{\wffallminarule}{\ensuremath{\phi_{\T{}(po)}^{all}}\xspace}
\newcommand{\wffallminaT}[1]{\ensuremath{\phi_{#1}^{all}}\xspace}
\newcommand{\wffallminaTrule}[1]{\ensuremath{\phi_{#1 (po)}^{all}}\xspace}
\newcommand{\wffallminaO}{\ensuremath{\phi_{\T^*}^{all}}\xspace}
\newcommand{\wffallminaruleO}{\ensuremath{\phi_{\T^*(po)}^{all}}\xspace}

\newcommand{\pinpointingwff}{\ensuremath{\Phi^{\goal}}\xspace}
\newcommand{\pinpointingwffall}{\ensuremath{\Phi^{all}}\xspace}

\newcommand{\hatT}{\ensuremath{\hat{\T}}\xspace}
\newcommand{\Tp}{\ensuremath{\ensuremath{\mathcal{T'}}\xspace}}

\newcommand{\elsat}{{\elp{}\textsc{SAT}}\xspace}

\newcommand{\el}{\ensuremath{\mathcal{EL}}\xspace}
\newcommand{\elh}{\ensuremath{\mathcal{ELH}}\xspace}
\newcommand{\elp}{\ensuremath{\mathcal{EL^{+}}}\xspace}
\newcommand{\elpp}{\ensuremath{\mathcal{EL^{++}}}\xspace}
\newcommand{\eloh}{\ensuremath{\mathcal{ELOH}}\xspace}
\newcommand{\eqdef}{\stackrel{def}{=}}



\newcommand{\nusmv}{{\sc NuSMV}}
\newcommand{\PROMELA}{{\sc PROMELA}}
\newcommand{\smv}{{\sc SMV}}

\newcommand{\obdds}{\text{OBDD}s\xspace}
\newcommand{\obdd}{\textrm{OBDD}\xspace}
\newcommand{\obddof}[1]{\textrm{OBDD}{\ensuremath{(#1)}}\xspace}
\newcommand{\Tobdds}{\text{\T-OBDD}s\xspace}
\newcommand{\Tobdd}{\textrm{\T-OBDD}\xspace}
\newcommand{\Tobddof}[1]{\textrm{\T-OBDD}{\ensuremath{(#1)}}\xspace}
\newcommand{\sdds}{\text{SDD}s\xspace}
\newcommand{\sdd}{\textrm{SDD}\xspace}
\newcommand{\sddof}[1]{\textrm{SDD}{\ensuremath{(#1)}}\xspace}
\newcommand{\Tsdds}{\text{\T-SDD}s\xspace}
\newcommand{\Tsdd}{\textrm{\T-SDD}\xspace}
\newcommand{\Tsddof}[1]{\textrm{\T-SDD}{\ensuremath{(#1)}}\xspace}

\newcommand{\bdds}{\text{DD}s\xspace}
\newcommand{\bdd}{\textrm{DD}\xspace}
\newcommand{\bddof}[1]{\textrm{DD}{\ensuremath{(#1)}}\xspace}
\newcommand{\Tbdds}{\text{\T-DD}s\xspace}
\newcommand{\Tbdd}{\textrm{\T-DD}\xspace}
\newcommand{\Tbddof}[1]{\textrm{\T-DD}{\ensuremath{(#1)}}\xspace}
\newcommand{\dis}{\vee}
\newcommand{\con}{\wedge}
\newcommand{\liff}{\leftrightarrow}
\newcommand{\true}{\top}
\newcommand{\false}{\bot}
\newcommand{\CTL}{\textsc{CTL}\xspace}
\newcommand{\LTL}{\textsc{LTL}\xspace}

\newcommand{\E}{{\bf E\/}}
\newcommand{\A}{{\bf A\/}}
\newcommand{\G}{{\bf G\/}}
\newcommand{\F}{{\bf F\/}}
\newcommand{\U}{{\bf U\/}}
\newcommand{\R}{{\bf R\/}}
\newcommand{\X}{{\bf X\/}}
\newcommand{\AG}{{\bf AG\/}}
\newcommand{\AF}{{\bf AF\/}}
\newcommand{\AU}{{\bf AU\/}}
\newcommand{\AR}{{\bf AR\/}}
\newcommand{\AX}{{\bf AX\/}}
\newcommand{\EG}{{\bf EG\/}}
\newcommand{\EF}{{\bf EF\/}}
\newcommand{\EU}{{\bf EU\/}}
\newcommand{\ER}{{\bf ER\/}}
\newcommand{\EX}{{\bf EX\/}}

\newcommand{\n}[2]{[[#1]]_{#2}}
\newcommand{\kil}[3]{\ _{#3}[[#1]]_{k}^{#2}}
\newcommand{\ki}[2]{\kil{#1}{#2}{}}
\renewcommand{\k}[1]{\kil{#1}{}{}}
\newcommand{\kl}[2]{\kil{#1}{}{#2}}

\newcommand{\fkil}[2]{\kil{f}{#1}{#2}}
\newcommand{\fki}[1]{\kil{f}{#1}{}}
\newcommand{\fk}{\kil{f}{}{}}
\newcommand{\fkl}[1]{\kil{f}{}{#1}}

\newcommand{\gkil}[2]{\kil{g}{#1}{#2}}
\newcommand{\gki}[1]{\kil{g}{#1}{}}
\newcommand{\gk}{\kil{g}{}{}}
\newcommand{\gkl}[1]{\kil{g}{}{#1}}

\newcommand{\Lkl}[1]{\ _{#1}{L}_{k}}
\newcommand{\Lkkll}[2]{\ _{#2}{L}_{#1}}
\newcommand{\Lkll}[1]{{L}_{k|#1}}
\newcommand{\Lk}{{L}_{k}}

\newcommand{\Mkil}[3]{\kil{M}{#1}{#2}{#3}}
\newcommand{\Mki}[2]{\kil{M}{#1}{#2}{}}
\newcommand{\Mk}[1]{\kil{M}{#1}{}{}}
\newcommand{\Mkl}[2]{\kil{M}{#1}{}{#2}}




\newcommand{\omtproblem}{\triple{\vi}{\costfs}{\bounds}\xspace}
\newcommand{\omt}{\ensuremath{\text{OMT}}\xspace}
\newcommand{\bomt}{\ensuremath{\text{BOMT}}\xspace}
\newcommand{\omtt}{\ensuremath{\text{OMT}({\T})}\xspace}
\newcommand{\bomtt}{\ensuremath{\text{BOMT}({\T})}\xspace}

\newcommand{\omtplus}[1]{\ensuremath{\text{OMT}({#1})}\xspace}
\newcommand{\omlarat}{\ensuremath{\text{OMT}(\larat)}\xspace}
\newcommand{\omlaint}{\ensuremath{\text{OMT}(\laint)}\xspace}
\newcommand{\omla}{\ensuremath{\text{OMT}(\la)}\xspace}
\newcommand{\omlaintplus}{\ensuremath{\text{OMT}(\laint\cup\T)}\xspace}
\newcommand{\omlaratplus}{\ensuremath{\text{OMT}(\larat\cup\T)}\xspace}
\newcommand{\omlaplus}{\ensuremath{\text{OMT}(\la\cup\T)}\xspace}
\newcommand{\omlaratint}{\ensuremath{\text{OMT}(\laratint)}\xspace}

\newcommand{\smtlaratplus}{\smttt{\larat\cup\T}\xspace}
\newcommand{\laratplus}{\ensuremath{\larat\cup\T}\xspace}

\newcommand{\costfs}{\ensuremath{\underline{costs}}\xspace}
\newcommand{\costfgen}[1]{\ensuremath{{cost}^{#1}}\xspace}
\newcommand{\costfi}{\costfgen{i}}
\newcommand{\costfone}{\costfgen{1}}
\newcommand{\costfM}{\costfgen{M}}

\newcommand{\costs}{\ensuremath{\underline{c}}\xspace}
\newcommand{\costgen}[1]{\ensuremath{{c}^{#1}}\xspace}
\newcommand{\costi}{\costgen{i}}
\newcommand{\costone}{\costgen{1}}
\newcommand{\costM}{\costgen{M}}

\newcommand{\vvgen}[2]{\ensuremath{{\sf v}^{#1}_{#2}}\xspace}
\newcommand{\vvi}[1]{\vvgen{i}{#1}}
\newcommand{\wvgen}[2]{\ensuremath{{\sf w}^{#1}_{#2}}\xspace}
\newcommand{\wvi}[1]{\wvgen{i}{#1}}
\newcommand{\xvgen}[2]{\ensuremath{{\sf x}^{#1}_{#2}}\xspace}
\newcommand{\xvi}[1]{\xvgen{i}{#1}}
\newcommand{\yvgen}[2]{\ensuremath{{\sf y}^{#1}_{#2}}\xspace}
\newcommand{\yvi}[1]{\yvgen{i}{#1}}

\newcommand{\lbgen}[2]{\ensuremath{{\sf lb}^{#1}_{#2}}\xspace}
\newcommand{\ubgen}[2]{\ensuremath{{\sf ub}^{#1}_{#2}}\xspace}
\newcommand{\lb}{\lbgen{}{}}
\newcommand{\ub}{\ubgen{}{}}
\newcommand{\lbi}{\lbgen{i}{}}
\newcommand{\ubi}{\ubgen{i}{}}
\newcommand{\lbone}{\lbgen{1}{}}
\newcommand{\ubone}{\ubgen{1}{}}
\newcommand{\lbM}{\lbgen{M}{}}
\newcommand{\ubM}{\ubgen{M}{}}
\newcommand{\lbigen}[1]{\lbgen{i}{(#1)}}
\newcommand{\ubigen}[1]{\ubgen{i}{(#1)}}
\newcommand{\lbimax}{\lbgen{i}{max}}
\newcommand{\ubimin}{\ubgen{i}{min}}
\newcommand{\lbonemax}{\lbgen{1}{max}}
\newcommand{\ubonemin}{\ubgen{1}{min}}

\newcommand{\pivot}{\ensuremath{\mathsf{pivot}}\xspace}
\newcommand{\trueval}{{\ensuremath{\mathsf{true}}}}
\newcommand{\falseval}{{\ensuremath{\mathsf{false}}}}
\newcommand{\tpredgen}[2]{{\ensuremath{\psi^{#1}_{#2}}}\xspace}
\newcommand{\tpredigen}[1]{\tpredgen{i}{#1}}
\newcommand{\tpredij}{\tpredigen{j}}
\newcommand{\predgen}[2]{{\ensuremath{A^{#1}_{#2}}}\xspace}
\newcommand{\predigen}[1]{\predgen{i}{#1}}
\newcommand{\predij}{\predigen{j}}
\newcommand{\predonej}{\predgen{1}{j}}

\newcommand{\Agen}[1]{\ensuremath{A_{#1}}\xspace}
\newcommand{\Aj}{\Agen{j}}
\newcommand{\aij}{\ensuremath{{\sf a}_{ij}}\xspace}
\newcommand{\bi}{\ensuremath{{\sf b}_{i}}\xspace}
\newcommand{\cj}{\ensuremath{{\sf c}_{j}}\xspace}
\newcommand{\ddj}{\ensuremath{{\sf d}_{j}}\xspace}
\newcommand{\lj}{\ensuremath{{l}_{j}}\xspace}

\newcommand{\cgen}[3]{{\ensuremath{{\sf c}^{#1}_{#2#3}}}\xspace}
\newcommand{\cigen}[2]{{\ensuremath{{\sf c}^{i}_{#1#2}}}\xspace}
\newcommand{\cijgen}[1]{{\ensuremath{{\sf c}^{i}_{j#1}}}\xspace}
\newcommand{\cijone}{{\ensuremath{{\sf c}^{i}_{j1}}}\xspace}
\newcommand{\cijtwo}{{\ensuremath{{\sf c}^{i}_{j2}}}\xspace}
\newcommand{\cij}{{\ensuremath{{\sf c}^{i}_{j}}}\xspace}
\newcommand{\conej}{{\ensuremath{{\sf c}^{1}_{j}}}\xspace}

\newcommand{\cc}{{\ensuremath{{\sf c}}}\xspace}

\newcommand{\bounds}{\ensuremath{\underline{\sf bounds}}\xspace}
\newcommand{\boundi}{\ensuremath{\langle\lbi,\ubi\rangle}\xspace}

\newcommand{\icost}{\ensuremath{{\sf IC}}\xspace}
\newcommand{\icostgen}[3]{\ensuremath{\icost(#3,#2,#1)}\xspace}
\newcommand{\icosti}[1]{\icostgen{\cigen{#1}{}}{#1}{\costgen{i}}}
\newcommand{\icostij}{\icostgen{\cij}{j}{\costgen{i}}}
\newcommand{\icostonej}{\icostgen{\cgen{1}{j}{}}{j}{\costgen{1}}}

\newcommand{\bcost}{\ensuremath{{\sf BC}}\xspace}
\newcommand{\cost}{\ensuremath{{cost}}\xspace}
\renewcommand{\costs}{\ensuremath{{costs}}\xspace}
\newcommand{\mincost}{\ensuremath{{mincost}}\xspace}
\newcommand{\bound}{\ensuremath{{bound}}\xspace}
\newcommand{\guess}{\ensuremath{{guess}}\xspace}
\newcommand{\oldguess}{\ensuremath{{oldguess}}\xspace}
\newcommand{\oldbound}{\ensuremath{{oldbound}}\xspace}
\newcommand{\maxboundi}{\ensuremath{{\sf bound}^{i}_{max}}\xspace}
\newcommand{\maxboundone}{\ensuremath{{\sf bound}^{1}_{max}}\xspace}
\newcommand{\nextsmallerbound}{\ensuremath{{\sf bound}_{next}}\xspace}
\newcommand{\noroom}{\ensuremath{{\sf NR}}\xspace}
\newcommand{\etastbe}{\ensuremath{{\sf MB}}\xspace}
\newcommand{\bleft}{\ensuremath{{lower}}\xspace}
\newcommand{\bright}{\ensuremath{{upper}}\xspace}
\newcommand{\pending}{\ensuremath{{\sf pending}}\xspace}

\newcommand{\tite}{\ensuremath{{ite}}\xspace}
\newcommand{\ite}{\ensuremath{\mathit{ITE}}\xspace}

\newcommand{\costof}[2]{\ensuremath{{\sf CostOf_{#1}}(#2)}\xspace}
\newcommand{\costiof}[1]{\costof{i}{#1}}
\newcommand{\mcostof}[2]{\ensuremath{{\sf MCostOf_{#1}}(#2)}\xspace}
\newcommand{\mcostiof}[1]{\mcostof{i}{#1}}


\newcommand{\C}{\ensuremath{\mathcal{C}}\xspace}
\newcommand{\Clemma}{\C-lemma\xspace}
\newcommand{\Csolver}{\TsolverGen{\C}}

\newcommand{\TC}{\ensuremath{\mathcal{T}\cup\mathcal{C}}\xspace}
\newcommand{\incsmttc}{\ensuremath{\text{IncrementalSMT}_{\TC}}\xspace}
\newcommand{\smttcost}{\ensuremath{\text{SMT}(\T)+cost}\xspace}
\newcommand{\smttc}{\ensuremath{\text{SMT}({\TC})}\xspace}
\newcommand{\smtc}{\ensuremath{\text{SMT}({\C})}\xspace}
\newcommand{\Tlaint}{\ensuremath{\T\cup\laint}\xspace}
\newcommand{\smttlaint}{\ensuremath{\text{SMT}(\T\cup\laint)}\xspace}

\newcommand{\optsmttc}{\ensuremath{\text{OptSMT}_{\TC}}\xspace}

\newcommand{\decproblem}{\ensuremath{\triple{\vi}{\cost}{\bound}}\xspace}
\newcommand{\optproblem}{\ensuremath{\pair{\vi}{\cost}}\xspace}

\newcommand{\vic}{\ensuremath{\vi_{\C}}\xspace}

\newcommand{\eqij}{\ensuremath{(x_i=x_j)}\xspace}
\newcommand{\neqij}{\ensuremath{\neg(x_i=x_j)}\xspace}
\newcommand{\dij}{\ensuremath{(x_i<x_j)}\xspace}
\newcommand{\dji}{\ensuremath{(x_i>x_j)}\xspace}

\newcommand{\etac}{\ensuremath{\eta_{\C}}\xspace}
\newcommand{\etacgen}[1]{\ensuremath{\eta^{#1}_{\C}}\xspace}
\newcommand{\etaci}{\etacgen{i}}
\newcommand{\etat}{\ensuremath{\eta_{\T}}\xspace}
\newcommand{\etab}{\ensuremath{\eta_{\calb}}\xspace}

\newcommand{\etalarat}{\ensuremath{\eta_{\larat}}\xspace}
\newcommand{\etaT}{\ensuremath{\eta_{\T}}\xspace}
\newcommand{\etabool}{\ensuremath{\eta_{\mathbb{B}}}\xspace}

\newcommand{\etae}{\ensuremath{\eta_{e}}\xspace}
\newcommand{\etad}{\ensuremath{\eta_{d}}\xspace}
\newcommand{\etai}{\ensuremath{\eta_{i}}\xspace}
\newcommand{\etaed}{\ensuremath{\eta_{ed}}\xspace}
\newcommand{\etaei}{\ensuremath{\eta_{ei}}\xspace}
\newcommand{\etaeid}{\ensuremath{\eta_{eid}}\xspace}

\newcommand{\etarelaxed}{\ensuremath{\eta_{rel}}\xspace}

\newcommand{\Jplus}{\ensuremath{J^{i+}}\xspace}
\newcommand{\Jminus}{\ensuremath{J^{i-}}\xspace}
\newcommand{\Kplus}{\ensuremath{K^{i+}}\xspace}
\newcommand{\Kminus}{\ensuremath{K^{i-}}\xspace}
\newcommand{\Kplusone}{\ensuremath{K^{1+}}\xspace}
\newcommand{\Kminusone}{\ensuremath{K^{1-}}\xspace}

\newcommand{\Ilaint}{\ensuremath{\cali_{\laint}}\xspace}
\newcommand{\IT}{\ensuremath{\cali_{\T}}\xspace}
\newcommand{\I}{\ensuremath{\cali}\xspace}

\newcommand{\mathsatC}{\mathsat}
\newcommand{\scip}{{\sc Scip}\xspace}
\newcommand{\bsolo}{{\sc Bsolo}\xspace}
\newcommand{\satforj}{{\sc Sat4j}\xspace}
\newcommand{\pbclasp}{{\sc PBClasp}\xspace}

\newcommand{\bl}{\phantom{0}}

\newcommand\mysout{\bgroup \markoverwith{{-}}\ULon}
\newcommand\nosout{\bgroup \markoverwith{{ }}\ULon}
\definecolor{mygray}{rgb}{0.90,0.90,0.90}
\definecolor{mywhite}{rgb}{1.00,1.00,1.00}
\def \gbox #1{\colorbox{mygray}{{#1}}}
\newcommand{\fal}[1]{\colorbox{mygray}{$\red{#1}$}}
\newcommand{\tru}[1]{\blue{#1}}

\newcommand{\mycite}[1]{{\footnotesize \textcolor{darkviolet}{#1}}}
\newcommand{\mcite}[1]{{\footnotesize \textcolor{darkviolet}{[#1]}}}

\newcommand{\etaeuf}{\ensuremath{\eta_{\euf}}\xspace}
\newcommand{\etalaint}{\ensuremath{\eta_{\laint}}\xspace}
\newcommand{\infrule}[3]{\ensuremath{\displaystyle\frac{#2}{#3}}}
\newcommand{\hyp}{\textsc{Hyp}\xspace}
\newcommand{\comb}{\textsc{Comb}\xspace}

\newcommand{\fakeinfrule}[3]{\ensuremath{\displaystyle\begin{array}{l}\phantom{#2}}\\{#3}\end{array}}


\newcommand{\currlb}{\ensuremath{\mathsf{l}}\xspace}
\newcommand{\currub}{\ensuremath{\mathsf{u}}\xspace}
\newcommand{\range}{\ensuremath{[\lb,\ub[}\xspace}
\newcommand{\currrange}{\ensuremath{[\currlb,\currub[}\xspace}
\newcommand{\lpivotrange}{\ensuremath{[\currlb,\pivot[}\xspace}
\newcommand{\rpivotrange}{\ensuremath{[\pivot,\currub[}\xspace}
\newcommand{\computepivot}{\ensuremath{{\sf ComputePivot}}\xspace}
\newcommand{\dopivoting}{\ensuremath{{\sf BinSearchMode()}}\xspace}

\newcommand{\ubliti}[1]{\ensuremath{(\cost < #1)}\xspace}
\newcommand{\lbliti}[1]{\ensuremath{\neg(\cost < #1)}\xspace}
\newcommand{\pivotatom}{\ensuremath{\mathsf{PIV}}\xspace}
\newcommand{\currlblit}{\ensuremath{\neg(\cost < \currlb)}\xspace}
\newcommand{\currublit}{\ensuremath{(\cost < \currub)}\xspace}
\newcommand{\currubliti}[1]{\ensuremath{(\cost < \currub_{#1})}\xspace}
\newcommand{\negcurrubliti}[1]{\ensuremath{\neg(\cost < \currub_{#1})}\xspace}

\newcommand{\minvalue}{\ensuremath{\mathsf{min}}\xspace}
\newcommand{\maxvalue}{\ensuremath{\mathsf{max}}\xspace}
\newcommand{\mvalue}{\ensuremath{\mathsf{m}}\xspace}
\renewcommand{\mincost}{\ensuremath{\mathsf{min}_\cost}\xspace}

\newcommand{\minimize}{\ensuremath{\mathsf{Minimize}}\xspace}
\newcommand{\isminimum}{\ensuremath{\mathsf{IsMinimum}}\xspace}
\newcommand{\incrementalsmt}{\ensuremath{\mathsf{SMT.IncrementalSolve}}\xspace}
\newcommand{\smtcoreextract}{\ensuremath{\mathsf{SMT.ExtractUnsatCore}}\xspace}
\newcommand{\incrementalsmtcore}{\ensuremath{\mathsf{SMT.IncrementalSolve\&ExtractCore}}\xspace}
\newcommand{\laqliterals}{\larat\ensuremath{\mathsf{LiteralsOf}}\xspace}

\newcommand{\optimathsat}{\textsc{OptiMathSAT}\xspace}
\newcommand{\optmathsat}{\textsc{OptiMathSAT}\xspace}
\newcommand{\symba}{\textsc{Symba}\xspace}


\newcommand{\dpll}{\textsc{DPLL}\xspace}

\newcommand{\zchaff}    {\textsc{Zchaff}\xspace}
\newcommand{\satelite}  {\textsc{SAT-Elite}\xspace}
\newcommand{\siege}     {\textsc{Siege}\xspace}
\newcommand{\berkmin}   {\textsc{BerkMin}\xspace}
\newcommand{\minisat}   {\textsc{MiniSat}\xspace}
\newcommand{\minisattwo}{\textsc{MiniSat2}\xspace}

\newcommand{\pickassign}{\ensuremath{\sf{pick\_assignment}}\xspace}
\newcommand{\picknewassign}{\ensuremath{\sf{pick\_new\_total\_assignment}}\xspace}
\newcommand{\picktotalassign}{\ensuremath{\sf{pick\_total\_assign}}\xspace}
\newcommand{\decidebranch}{\ensuremath{\sf{decide\_new\_branch}}\xspace}
\newcommand{\decide}{\ensuremath{\sf{decide}}\xspace}
\newcommand{\blevel}{\ensuremath{\sf{blevel}}\xspace}
\newcommand{\analyzeconflict}{\ensuremath{\sf{analyze\_conflict}}\xspace}
\newcommand{\analyzededuction}{\ensuremath{\sf{analyze\_deduction}}\xspace}
\newcommand{\backtrack}{\ensuremath{\sf{backtrack}}\xspace}
\newcommand{\bcp}{\ensuremath{\sf{boolean\_constraint\_propagation}}\xspace}

\newcommand{\satres}{\textsc{sat}\xspace}
\newcommand{\unsatres}{\textsc{unsat}\xspace}
\newcommand{\unknownres}{\textsc{unknown}\xspace}
\newcommand{\conflres}{\ensuremath{\mathsf{conflict}}\xspace}

\newcommand{\walksat}{WalkSAT\xspace}
\newcommand{\basicwalksmt}{\textsc{Basic-WalkSMT}\xspace}
\newcommand{\walksmt}{\textsc{WalkSMT}\xspace}

\newcommand{\tpreprocess}{\T{}{\sc -preprocess}\xspace}
\newcommand{\initialtruthassignment}{{\sc InitialTruthAssignment}\xspace}
\newcommand{\chooseunsatisfiedclause}{{\sc ChooseUnsatisfiedClause}\xspace}
\newcommand{\nexttruthassignment}{{\sc NextTruthAssignment}\xspace}


\newcommand{\proptofol}{\ensuremath{{\cal B}2{\cal T}}\xspace}
\newcommand{\foltoprop}{\ensuremath{{\cal T}2{\cal B}}\xspace}
\newcommand{\btot}{\proptofol}
\newcommand{\ttob}{\foltoprop}

\newcommand{\vip}{\ensuremath{\varphi^p}\xspace}
\newcommand{\etap}{\ensuremath{\eta^p}\xspace}
\newcommand{\etaone}{\ensuremath{\eta_1}\xspace}
\newcommand{\etatwo}{\ensuremath{\eta_2}\xspace}
\newcommand{\etaonetwo}{\ensuremath{\eta_1 \cup \eta_2}\xspace}
\newcommand{\taup}{\ensuremath{\tau^p}\xspace}
\newcommand{\rhop}{\ensuremath{\rho^p}\xspace}
\newcommand{\cp}{\ensuremath{c^p}\xspace}

\newcommand{\atoms}[1]{\ensuremath{Atoms(#1)}\xspace}

\newcommand{\B}{\ensuremath{\mathcal{B}}\xspace}
\newcommand{\T}{\ensuremath{\mathcal{T}}\xspace}
\newcommand{\Tone}{\ensuremath{\T_1}\xspace}
\newcommand{\Ttwo}{\ensuremath{\T_2}\xspace}
\newcommand{\Tonetwo}{\ensuremath{\Tone\cup \Ttwo}\xspace}
\newcommand{\smt}{SMT\xspace}
\newcommand{\smtt}{\ensuremath{\text{SMT}(\T)}\xspace}
\newcommand{\smttt}[1]{\ensuremath{\text{SMT}(#1)}\xspace}
\newcommand{\smttonettwo}{\ensuremath{\text{SMT}(\tonetwo)}\xspace}
\newcommand{\Ti}{\ensuremath{\T_i}\xspace}
\newcommand{\tmany}{\ensuremath{\bigcup_i \T_i}\xspace}
\newcommand{\smtmany}{\ensuremath{\mathit{SMT}(\bigcup_i \T_i)}\xspace}

\newcommand{\utvpi}{\ensuremath{\mathcal{UTVPI}}\xspace}
\newcommand{\utvpiint}{\ensuremath{\mathcal{UTVPI(\mathbb{Z})}}\xspace}
\newcommand{\utvpirat}{\ensuremath{\mathcal{UTVPI(\mathbb{Q})}}\xspace}
\newcommand{\bool}{\ensuremath{\mathcal{BOOL}}\xspace}
\newcommand{\euf}{\ensuremath{\mathcal{EUF}}\xspace}
\newcommand{\sk}{\ensuremath{\mathcal{SK}}\xspace}
\newcommand{\eq}{\ensuremath{\mathcal{E}}\xspace}
\newcommand{\dl}{\ensuremath{\mathcal{DL}}\xspace}
\newcommand{\dlrat}{\ensuremath{\mathcal{DL(\mathbb{Q})}}\xspace}
\newcommand{\dlint}{\ensuremath{\mathcal{DL(\mathbb{Z})}}\xspace}
\newcommand{\la}{\ensuremath{\mathcal{LA}}\xspace}
\newcommand{\larat}{\ensuremath{\mathcal{LA}(\mathbb{Q})}\xspace}
\newcommand{\laint}{\ensuremath{\mathcal{LA}(\mathbb{Z})}\xspace}
\newcommand{\laratint}{\ensuremath{\mathcal{LA}(\mathbb{Q}\mathbb{Z})}\xspace}
\renewcommand{\la}{\ensuremath{\mathcal{LA}}\xspace}
\renewcommand{\larat}{\ensuremath{\mathcal{LRA}}\xspace}
\renewcommand{\laint}{\ensuremath{\mathcal{LIA}}\xspace}
\renewcommand{\laratint}{\ensuremath{\mathcal{LIRA}}\xspace}

\newcommand{\fl}{\ensuremath{\mathcal{FP}}\xspace}
\newcommand{\nla}{\ensuremath{\mathcal{NLA}}\xspace}
\newcommand{\nlarat}{\ensuremath{\mathcal{NRA}}\xspace}
\newcommand{\nlaint}{\ensuremath{\mathcal{NIA}}\xspace}
\newcommand{\nlaratint}{\ensuremath{\mathcal{NIRA}}\xspace}
\newcommand{\bv}{\ensuremath{\mathcal{BV}}\xspace}
\newcommand{\mem}{\ensuremath{\mathcal{AR}}\xspace}
\newcommand{\lists}{\ensuremath{\mathcal{LI}}\xspace}
\newcommand{\tonetwo}{\ensuremath{\T_1\cup\T_2}\xspace}

\newcommand{\smteuf}{\smttt{\euf}}
\newcommand{\smtdl}{\smttt{\dl}}
\newcommand{\smtdlrat}{\smttt{\dlrat}}
\newcommand{\smtdlint}{\smttt{\dlint}}
\newcommand{\smtutvpi}{\smttt{\utvpi}}
\newcommand{\smtutvpirat}{\smttt{\utvpirat}}
\newcommand{\smtutvpiint}{\smttt{\utvpiint}}
\newcommand{\smtla}{\smttt{\la}}
\newcommand{\smtT}{\smttt{\T}}
\newcommand{\smtlarat}{\smttt{\larat}}
\newcommand{\smtlaint}{\smttt{\laint}}
\newcommand{\smtbv}{\smttt{\bv}}
\newcommand{\smtmem}{\smttt{\mem}}

\newcommand{\genmodels}[1]{\models_{#1}}
\newcommand{\Tmodels}{\models_{\T}}
\newcommand{\laratmodels}{\models_{\larat}}
\newcommand{\laintmodels}{\models_{\laint}}
\newcommand{\lamodels}{\models_{\la}}
\newcommand{\dlmodels}{\models_{\dl}}
\newcommand{\dlintmodels}{\models_{\dlint}}
\newcommand{\dlratmodels}{\models_{\dlrat}}
\newcommand{\pmodels}{\models_p}

\newcommand{\TsolverGen}[1]{\ensuremath{{#1}\textit{-solver}}\xspace}
\newcommand{\TsolversGen}[1]{\ensuremath{{#1}\textit{-solvers}}\xspace}
\newcommand{\Tsolver}{\TsolverGen{\T}}
\newcommand{\Tsolvers}{\TsolversGen{\T}}
\newcommand{\Psolver}{\TsolverGen{\mathsf{Bool}}}
\newcommand{\TTsolver}[1]{\TsolverGen{\ensuremath{#1}}}
\newcommand{\Tonesolver}{\TsolverGen{\ensuremath{\Tone}}}
\newcommand{\Ttwosolver}{\TsolverGen{\ensuremath{\Ttwo}}}
\newcommand{\Tonetwosolver}{\TsolverGen{\ensuremath{\Tone\cup \Ttwo}}}
\newcommand{\Tisolver}{\TsolverGen{\ensuremath{\Ti}}}
\newcommand{\Tisolvers}{\TsolversGen{\ensuremath{\Ti}}}
\newcommand{\TdeduceGen}[1]{\ensuremath{{#1}\textit{-deduce}}\xspace}
\newcommand{\Tonededuce}{\TdeduceGen{\ensuremath{\Tone}}}
\newcommand{\Ttwodeduce}{\TdeduceGen{\ensuremath{\Ttwo}}}

\newcommand{\Tlemma}{\T-lemma\xspace}
\newcommand{\Tlemmas}{\T-lemmas\xspace}
\newcommand{\Tilemma}{\Ti-lemma\xspace}
\newcommand{\Tilemmas}{\Ti-lemmas\xspace}

\newcommand{\laratsolver}{\larat-\ensuremath{\mathsf{Solver}}}
\newcommand{\laratsolvers}{\larat-\ensuremath{\mathsf{Solvers}}}
\newcommand{\laintsolver}{\laint-\ensuremath{\mathsf{Solver}}}
\newcommand{\laintsolvers}{\laint-\ensuremath{\mathsf{Solvers}}}
\newcommand{\lasolver}{\la-\ensuremath{\mathsf{Solver}}}
\newcommand{\lasolvers}{\la-\ensuremath{\mathsf{Solvers}}}


\newcommand{\argolib}{\textsc{ArgoLib}\xspace}
\newcommand{\barcelogic}{\textsc{Barcelogic}\xspace}
\newcommand{\blast}{\textsc{Blast}\xspace}
\newcommand{\clp}{\textsc{clp-prover}\xspace}
\newcommand{\csisat}{\textsc{CSIsat}\xspace}
\newcommand{\cvcthree}{\textsc{CVC3}\xspace}
\newcommand{\foci}{\textsc{Foci}\xspace}
\newcommand{\haRVey}{\textsc{haRVey}\xspace}
\newcommand{\intjain}{\textsc{INT2}\xspace}
\newcommand{\lifter}{\textsc{Lifter}\xspace}
\newcommand{\mathsat}{\textsc{MathSAT}\xspace}
\newcommand{\verifun} {\textsc{Verifun}\xspace}
\newcommand{\yices}{\textsc{Yices}\xspace}
\newcommand{\zapato}  {\textsc{Zapato}\xspace}
\newcommand{\zap}{\textsc{Zap}\xspace}
\newcommand{\zthree}{\textsc{Z3}\xspace}

\newcommand{\microformal}{\textsc{MicroFormal}\xspace}

\newcommand{\mathsatfour}{\textsc{MathSAT4}\xspace}
\newcommand{\mathsatfive}{\textsc{MathSAT5}\xspace}
\newcommand{\mathsatfivePrep}{\textsc{MathSAT5}$_{\textsc{preprocessing}}$\xspace}
\newcommand{\mathsatfiveCleaneling}{\textsc{MathSAT5}$_{\cleaneling}$\xspace}
\newcommand{\mathsatfiveMinisat}{\textsc{MathSAT5}$_{\minisat}$\xspace}

\newcommand{\myeij}{\ensuremath{e_{ij}}\xspace}
\newcommand{\myeijs}{\ensuremath{e_{ij}}s\xspace}
\newcommand{\mynegeij}{\ensuremath{\neg e_{ij}}\xspace}
\newcommand{\eij}{\ensuremath{v_i=v_j}\xspace}
\newcommand{\eik}{\ensuremath{v_i=v_k}\xspace}
\newcommand{\ejk}{\ensuremath{v_j=v_k}\xspace}
\newcommand{\diseij}{\ensuremath{\vee(\eij)}\xspace}

\newcommand{\NO}{\ensuremath{\mathsf{no}(T_1,T_2)}\xspace}
\newcommand{\Bool}{\ensuremath{\mathsf{Bool}}\xspace}
\newcommand{\BoolGen}[1]{\ensuremath{\mathsf{Bool+}{#1}}\xspace}
\newcommand{\BoolT}{\BoolGen{T}}
\newcommand{\BoolTonetwo}{\BoolGen{T_1+T_2}}
\newcommand{\BoolTi}{\BoolGen{{T_i}}}
\newcommand{\BoolNO}{\BoolGen{\NO}}
\newcommand{\DelBoolNO}{\ensuremath{{\mathsf{Bool+}}T_1{\mathsf{+}}T_2}\xspace}

\newcommand{\TsatisfiableGen}[1]{\ensuremath{{#1}\textit{-satisfiable}}\xspace}
\newcommand{\TsatGen}[1]{\ensuremath{{#1}\textit{-sat}}\xspace}
\newcommand{\TunsatGen}[1]{\ensuremath{{#1}\textit{-unsat}}\xspace}
\newcommand{\Tonesat}{\TsatGen{\ensuremath{T_1}}}
\newcommand{\Ttwosat}{\TsatGen{\ensuremath{T_2}}}
\newcommand{\Toneunsat}{\TunsatGen{\ensuremath{T_1}}}
\newcommand{\Ttwounsat}{\TunsatGen{\ensuremath{T_2}}}
\newcommand{\Tsatisfiable}{\TsatisfiableGen{T}}
\newcommand{\Psatisfiable}{\TsatisfiableGen{\mathsf{Bool}}}
\newcommand{\Tonesatisfiable}{\TsatisfiableGen{\ensuremath{T_1}}}
\newcommand{\Ttwosatisfiable}{\TsatisfiableGen{\ensuremath{T_2}}}
\newcommand{\Tisatisfiable}{\TsatisfiableGen{\ensuremath{T_i}}}

\setlength{\marginparwidth}{2.5cm}
\renewcommand{\TODO}[1]{\todo[inline,color=green!40]{{\small{TODO: #1}}}}
\newcommand{\MARGTODO}[1]{\todo[color=green!40]{{\small{#1}}}}
\newcommand{\NOTE}[1]{\todo[inline,color=orange!40]{{\small{#1}}}}
\renewcommand{\RSTODO}[1]{\todo[inline,color=green!40]{{\small{RS TODO: #1}}}}
\renewcommand{\RSNOTE}[1]{\todo[inline,color=green!40]{{\small{RS: #1}}}}
\newcommand{\RSSIDENOTE}[1]{\todo[color=green!40]{{\tiny{RS: #1}}}}
\newcommand{\GMSIDENOTE}[1]{\todo[color=orange!40]{{\small{GM: #1}}}}
\newcommand{\GMNOTE}[1]
{\todo[inline,color=orange!40]{{\small{GM: #1}}}}
\newcommand{\GMCHANGE}[1]{\textcolor{blue}{#1}}
\newenvironment{gmchange}{\color{blue}}{\normalcolor}
\newcommand{\GMTODO}[1]
{\todo[inline,color=orange!40]{{\small{GM TODO: #1}}}}
\newcommand{\REBUTTAL}[1]{\textcolor{blue}{#1}}


\newcommand{\M}{\ensuremath{\mathcal{M}}\xspace}
\renewcommand{\B}{\ensuremath{\mathcal{B}}\xspace}
\newcommand{\N}{\ensuremath{\mathcal{N}}\xspace}
\renewcommand{\minimize}{\textsf{T-Solver.Minimize}\xspace}
\newcommand{\minimizestar}{\ensuremath{\minimize\textsf{Approx}}\xspace}
\newcommand{\tpop}{\textsc{T-pop}\xspace}
\newcommand{\proposelit}{\textsf{T-Solver.ProposeLiteralToDrop}\xspace}
\newcommand{\vi}{\ensuremath{\varphi}}
\newcommand{\obj}{\ensuremath{\mathsf{cost}}}
\newcommand{\muprime}{\ensuremath{\mu^\prime}}
\newcommand{\mupprime}{\ensuremath{\mu^{\prime\prime}}}
\renewcommand{\smt}{\ensuremath{\text{SMT}}\xspace}
\newcommand{\viog}{\ensuremath{\vi}\xspace}
\newcommand{\viabs}{\ensuremath{\hat\vi}\xspace}
\newcommand{\muog}{\ensuremath{\mu}\xspace}
\newcommand{\Mabs}{\ensuremath{\hat\M}\xspace}
\newcommand{\Mabsopt}{\ensuremath{\hat\M^\prime}\xspace}
\newcommand{\Mprime}{\ensuremath{\M^\prime}\xspace}
\newcommand{\residual}[2]{\ensuremath{#1|_{#2}}\xspace}
\newcommand{\minimizeassignment}{\textsc{reduce-assignment}\xspace}
\newcommand{\omtminimizeassignment}{\textsc{OMT-reduce-assignment}\xspace}
\newcommand{\omtminimizeassignmentguided}{\textsc{OMT-reduce-assignment-guided}\xspace}

\colorlet{plane}{gray}
\newcommand{\planeopacity}{0.2}
\newcommand{\laratplusa}{\ensuremath{\larat\cup\mem}\xspace}
\newcommand{\omlaratplusa}{\ensuremath{\text{OMT}(\laratplusa)}\xspace}
\newcommand{\omlaratintplus}{\ensuremath{\text{OMT}(\laratint\cup\T)}\xspace}

%
\title{Exploiting Partial-Assignment Enumeration\\in Optimization Modulo Theories}
%
%
\author{Gabriele Masina\inst{1}\orcidID{0000-0001-8842-4913} \and
    Roberto Sebastiani\inst{1}\orcidID{0000-0002-0989-6101}}
\authorrunning{G. Masina and R. Sebastiani}
%
\institute{DISI, University of Trento, Trento, Italy
    \email{\{gabriele.masina,roberto.sebastiani\}@unitn.it}}
\maketitle              
\begin{abstract}

Optimization Modulo Theories (OMT) extends Satisfiability Modulo Theories (SMT) with the task of optimizing some objective function(s). 
In  OMT solvers,
a CDCL-based SMT solver enumerates theory-satisfiable total truth assignments, and 
a theory-specific  procedure finds an optimum model for each of them;
the current optimum is then used to tighten the search space for the next assignments, until no better solution is found.

In this paper, we analyze the role of truth-assignment enumeration in OMT. First, we spotlight that the enumeration of \emph{total} truth assignments is suboptimal, since they may over-restrict the search space for the optimization procedure, whereas using \emph{partial} truth assignments instead can improve the effectiveness of the optimization. 
Second, we propose some assignment-reduction techniques for exploiting \emph{partial-assignment enumeration} within the OMT context.
We implemented these techniques in the \optimathsat{} solver, and we conducted an experimental evaluation on \omt{} benchmarks.
The results confirm the improvement in both the efficiency of optimal solving and the quality of the obtained solutions for anytime solving.
    \keywords{Optimization Modulo Theories  \and Enumeration \and Partial Assignments.}
\end{abstract}
\section{Introduction}

Satisfiability Modulo Theories (SMT) is the problem of deciding the
satisfiability of a logical formula w.r.t.\ some background theory, such as
linear and nonlinear arithmetic, bit-vectors, arrays, or uninterpreted
functions~\cite{barrettSatisfiabilityModuloTheories2021}.
Many SMT-encodable problems also require the capability of finding models that
are optimal w.r.t.\ some objective functions. These problems are grouped under
the term Optimization Modulo Theories
(OMT)~\cite{nieuwenhuisSATModuloTheories2006,sebastianiOptimizationSMTLAQ2012,bjornerNZOptimizingSMT2015}.
%
OMT has been successfully applied to a wide range of problems, such as
verification of timed and hybrid
systems~\cite{sebastianiOptimizationSMTLAQ2012,henryHowComputeWorstcase2014},
numeric~\cite{leofanteOptimalPlanningModulo2021} and temporal
planning~\cite{panjkovicExpressiveOptimalTemporal2023,panjkovicAbstractActionScheduling2024},
optimal scheduling~\cite{bofillEfficientSMTApproach2017}, constrained goal
modelling~\cite{nguyenMultiobjectiveReasoningConstrained2018}, hybrid machine
learning~\cite{tesoStructuredLearningModulo2017}, GAS optimization for smart
contracts~\cite{albertGASOLGasAnalysis2020}, and optimum encodings for quantum
annealing~\cite{bianSolvingSATMaxSAT2020,dingEffectivePrimeFactorization2024},
establishing OMT solvers as powerful tools for solving complex constraint
optimization problems in various domains.



\paragraph{OMT solving.}%
\label{sec:related-work-omt}
%
A general OMT-solving
strategy~\cite{nieuwenhuisSATModuloTheories2006,sebastianiOptimizationSMTLAQ2012,sebastianiOptimizationModuloTheories2015}
consists in performing a sequence of incremental SMT calls, progressively
tightening the range of values for the objective function.
Specifically, an \smt{} solver is used to enumerate \T-satisfiable truth
assignments that propositionally satisfy the problem formula $\vi$. For each
such truth assignment, a \T-optimizer finds a \T-model of optimum cost within
it. A constraint is then added to the formula to tighten the upper bound for
the cost of the optimum model, and the search continues until the formula is
found unsatisfiable. Besides optimal solving, an important feature of \omt{}
solvers is the ability to provide the user with a good-enough solution within a
given time budget. This capability, known as \emph{anytime} OMT solving, is
especially valuable in industrial applications where finding the optimum
solution may be computationally impractical, and it is rather more important to
obtain high-quality solutions quickly.

\omt{} techniques have been developed for \larat{}~\cite{bjornerNZOptimizingSMT2015,sebastianiOptimizationModuloTheories2015}, \laratint{}~\cite{bjornerNZOptimizingSMT2015,sebastianiPushingEnvelopeOptimization2015}, \nlarat~\cite{bigarellaOptimizationModuloNonlinear2021}, \nlaint~\cite{bigarellaOptimizationModuloNonlinear2021}, \bv~\cite{nadelBitVectorOptimization2016,trentinOptimizationModuloTheories2021}, and \fl~\cite{trentinOptimizationModuloTheories2021}.
Also, \omt{} has been extended to deal with multiple objectives including lexicographic \omt{}~\cite{bjornerNZOptimizingSMT2015,sebastianiPushingEnvelopeOptimization2015}, boxed \omt{}~\cite{liSymbolicOptimizationSMT2014,bjornerNZOptimizingSMT2015,sebastianiPushingEnvelopeOptimization2015}, min-max \omt{}~\cite{sebastianiOptiMathSATToolOptimization2020}, and Pareto \omt{}~\cite{bjornerNZOptimizingSMT2015}. 
Recently, a Generalized OMT calculus has been proposed, extending the
definition to objectives over partially ordered
sets~\cite{tsiskaridzeGeneralizedOptimizationModulo2024}.

\paragraph{Partial assignments enumeration in SMT.}%
\label{sec:related-work-partial-assignments}

The problem of truth assignment enumeration has been studied in recent years,
mainly in the context of SAT and \smt{} enumeration (AllSAT and AllSMT).
Typically, enumeration
algorithms~\cite{lahiriSMTTechniquesFast2006,friedAllSATCombinationalCircuits2023,friedEntailingGeneralizationBoosts2024,spallittaDisjointPartialEnumeration2024,spallittaDisjointProjectedEnumeration2025}
rely their efficiency on the enumeration of partial assignments to reduce both
the number of enumerated assignments and the computational time by up to an
exponential factor.
%
Several techniques have been proposed to find short satisfying partial
assignments starting from a total assignment, trading off efficiency for
effectiveness
(e.g.,~\cite{raviMinimalAssignmentsBounded2004,morgadoGoodLearningImplicit2005,todaImplementingEfficientAll2016}).
Also, the impact of CNF-ization on the effectiveness of partial assignment
reduction has been recently studied
in~\cite{masinaCNFConversionDisjoint2023,spallittaEnhancingSMTbasedWeighted2024}.





\paragraph{Contributions.}
In this paper, we study the applicability of enumeration-based techniques to
OMT solving, and, in particular, the usage of partial truth assignment
reduction to improve the effectiveness and efficiency of OMT solving. First, we
notice that OMT solvers typically invoke the \T-optimizer on total truth
assignments, and we spotlight how this can be suboptimal in many cases. Second,
we propose some ways to exploit partial truth assignments in OMT solving,
tailoring existing techniques to the OMT context. We discuss the general idea
for an arbitrary theory, and describe an implementation for the specific case
of linear arithmetic over \larat{} and \laratint{}, possibly combined with
other theories. We implemented these strategies in the \omt{} solver
\optimathsat{}~\cite{sebastianiOptiMathSATToolOptimization2020}, and we show
through an empirical evaluation over \omlarat{}, \omlaratint{}, and
\omlaratplusa{} benchmarks that they improve both the efficiency of optimal
solving and the quality of obtained solutions for anytime solving. 

\paragraph{Related work.}
The idea of using partial assignments in \omlaratint{} has been previously
considered
in~\cite{maSolvingGeneralizedOptimization2012,bjornerNZMaximalSatisfaction2014}.
In~\cite{bjornerNZMaximalSatisfaction2014}, the authors mention that in lazy
OMT-solving, the truth assignments should preferably be prime implicants. This
approach, however, is theory-blind; our reduction techniques, instead,
specifically target theory-literals involved in optimization, which proves
crucial for the solver efficiency. A similar idea was proposed
in~\cite{maSolvingGeneralizedOptimization2012}, where the truth assignments are
reduced to prime implicants before invoking the \T-solver and \T-minimizer.
Though similar in flavour to our work, there are some key differences:
\begin{enumerate*}[label=(\alph*)]
    \item their enumeration algorithm only adds clauses blocking the minimized truth
          assignment, whereas modern \omt{} solvers add cost bounds $(\obj{} < \ub)$ to
          prune the search space, which has been shown to be much more effective;
    \item we also propose a cost-guided technique, which we show to perform much better
          in practice.
\end{enumerate*}

\paragraph{Organization.}
The rest of the paper is organized as follows. In \sref{sec:background}, we
provide the necessary background on SMT and OMT solving. In
\sref{sec:analysis}, we analyze the role of total and partial truth assignments
in OMT solving. In \sref{sec:approach}, we propose two strategies to exploit
partial truth assignments in OMT solving. In \sref{sec:experiments}, we present
an experimental evaluation of the proposed strategies over
\omlarat{}, \omlaratint{}, and \omlaratplusa{} benchmarks. Finally,
in \sref{sec:conclusions}, we conclude the paper and discuss future work.

\section{Background}%
\label{sec:background}
\paragraph{Notation and terminology.}
We assume the standard setting with quantifier-free first-order formulas, and
the standard notions of theory, satisfiability, logical consequence. We assume
the reader is familiar with these notions and with the lazy CDCL-based
SMT-solving approach, and refer
to~\cite{barrettSatisfiabilityModuloTheories2021} for a comprehensive
introduction to SMT.

In this paper, we denote SMT formulas by $\vi$, theories by $\T$, variables by
$x,y$, atoms by $\alpha$, truth assignments by $\mu,\eta$, and models by $\M$;
all symbols possibly with subscripts or superscripts. We denote by \atoms{\vi}
the set of atoms occurring in a formula \vi.

\subsection{Satisfiability Modulo Theories}%
\label{sec:bg:smt}
Given a first-order theory \T, a \T-atom is any atomic formula built over the signature of \T. A \T-literal is a \T-atom or its negation. A \T-formula is either a \T-literal or a combination of formulas by means of standard Boolean operators. From now on, we assume every formula is in Conjunctive Normal Form
(CNF), i.e., it is a conjunction ($\wedge$) of clauses, where each clause is a
disjunction ($\vee$) of literals. (If it is not, then it can be easily
converted into CNF by applying the standard
transformations~\cite{tseitinComplexityDerivationPropositional1983,plaistedStructurepreservingClauseForm1986}).

Satisfiability Modulo Theories (SMT) is the problem of deciding the
satisfiability of a first-order formula w.r.t\ some first-order theory \T, or
combination of first-order theories $\T \cup \T^\prime$. A formula is
\T-satisfiable if it is satisfiable in a model of \T (a \T-model). Popular
theories include linear and nonlinear arithmetic over the reals or integers
(\larat, \nlarat, \laratint, and \nlaratint), bit-vectors (\bv), and
floating-point (\fl).

\paragraph{Lazy SMT-solving.}

Given a formula $\vi$ with $\atoms{\vi}\defas\{\alpha_1,\ldots,\alpha_n\}$, a
truth assignment $\mu : \atoms{\vi} \to \{\top, \bot\}$ maps atoms in $\vi$ to
truth values. A partial truth assignment is a partial mapping, and a total
truth assignment is a total mapping. We represent a truth assignment $\mu$ also
as a conjunction of literals $\bigwedge_{\mu(\alpha_i)=\top}\alpha_i \wedge
  \bigwedge_{\mu(\alpha_i)=\bot}\neg\alpha_i$.
%
We say that $\mu$ \emph{propositionally satisfies} $\vi$ iff $\mu$ satisfies
all clauses in $\vi$.

The CDCL(\T) algorithm~\cite{marques-silvaConflictDrivenClauseLearning2021} is
based on the so-called lazy approach to SMT (see
e.g.,~\cite{sebastianiLazySatisfiabilityModulo2007,barrettSatisfiabilityModuloTheories2021}),
which exploits the fact that a \T-formula $\vi$ is \T-satisfiable iff there
exists a truth assignment $\mu$ that propositionally satisfies $\vi$ and $\mu$
is \T-satisfiable.
It combines a CDCL-based SAT-solver with a \T-specialized decision procedure
called \T-solver to decide the consistency of a set of \T-literals. Whenever
the SAT-solver finds a truth assignment $\mu$ propositionally satisfying $\vi$,
it invokes the \T-solver to check the \T-satisfiability of $\mu$. If $\mu$ is
\T-satisfiable, then the \T-solver returns a model $\M$, that is also a model
of $\vi$. Otherwise, the \T-solver returns a subset of $\mu$ that causes the
\T-unsatisfiability, which is learned by the SAT-solver and used in subsequent
iterations to prune the search space.
%


To maximize efficiency, most \T-solvers can be called incrementally via a
stack-based interface, keeping the status of the search between calls. E.g., an
efficient, incremental \larat-solver, can be built on a variant of the Simplex
algorithm designed to be integrated within a lazy SMT
framework~\cite{dutertreFastLinearArithmeticSolver2006}. The combination of
theories can be handled efficiently by delayed theory
combination~\cite{bozzanoEfficientTheoryCombination2006}.

Another important feature of CDCL-based SMT solvers is that they provide a
stack-based incremental interface, allowing to push and pop clauses and
incrementally check the satisfiability of the formula conjoined with the pushed
clauses, maintaining most of the learned information between calls.

\subsection{Optimization Modulo Theories}%
\label{sec:bg:omt}
Let \T{} be a theory admitting some total order relation ``$\leq$'' over its
domain, let \vi{} be a \T-formula, and let \obj{} be a \T-term which we call
\emph{objective function}. \emph{Optimization Modulo Theories} (\omtt) is the problem
of finding a model for \vi{} that makes the value of \obj{} minimum according
to the order given by $\leq$ (maximization is
dual)~\cite{sebastianiOptimizationSMTLAQ2012,bigarellaOptimizationModuloNonlinear2021}.
To simplify the presentation, we focus on minimization, but the same concepts
apply to maximization as well. 
Notice that, in general, \vi{} can be built on a combination of \T with other
theories ($\omt(\T\cup\T^\prime)$), and the same concepts apply to such
cases~\cite{sebastianiOptimizationSMTLAQ2012,sebastianiOptimizationModuloTheories2015}.

\begin{example}%
  \label{ex:smt}
  Consider the \larat-formula on the rational variables $x, y$:
  \begin{equation}%
    \label{eq:smt}
    \begin{array}{ll}
      \vi\defas & ((2x-3y\leq 6)\vee(x\leq 4))\wedge     \\
                & ((y\leq 2)\vee(y\leq-3x+9)\vee(x<-2)).
    \end{array}
  \end{equation}
  \vi{} is \larat-satisfiable, e.g., the \larat-model $\M\defas\set{x\mapsto{} 3, y\mapsto{} 0}$ satisfies $\vi$.

  Consider the \omlarat{} problem \pair{\vi}{\obj} where $\vi$ is the
  \larat-formula in~\eqref{eq:smt}, and $\obj\defas-2x$. Then the model
  $\M\defas\set{x\mapsto{} 3, y\mapsto{} 0}$ has $\obj=-6$. A better model of
  \vi{} is, e.g., $\Mprime\defas\set{x\mapsto{} 6, y\mapsto{} 2}$, that has
  $\obj=-12$. This model is also the model of \vi{} with minimum cost.
\end{example}

\paragraph{Lazy OMT solving.}%
\label{sec:omt-solving}

A general optimization strategy implemented by state-of-the-art \omt{} solvers,
and typically used for \omlarat{} and \omlaratint{}, is the so-called
\emph{linear-search}
strategy~\cite{nieuwenhuisSATModuloTheories2006,sebastianiOptimizationSMTLAQ2012,sebastianiOptimizationModuloTheories2015}.
It consists in solving a sequence of \smt{} problems where the space of
feasible solutions is progressively tightened by learning unit clauses in the
form $(\obj < \ub)$, \ub{} being the currently-known upper bound for \obj. At
each iteration, the solver can either find a model $\M$ whose value of \obj{}
is smaller than \ub{}, or detect the unsatisfiability of the current formula.
In the first case, the solver invokes a \T-specific procedure, called
\emph{\T-minimizer}, to find an optimum model $\Mprime$ within the truth
assignment induced by \M. Then, $\ub$ is set to $\Mprime(\obj)$, and the search
continues. In the second case, the formula has no models with \obj{} lower than
\ub{}, and the search terminates as the last model found is optimum.

Alternatively, the solver could also follow a \emph{binary-search}
strategy~\cite{sebastianiOptimizationSMTLAQ2012}. In this case, a lower and
upper bound \lb{} and \ub{} are kept s.t.\ the optimum model lies in the
interval $\mathopen(\lb{},\ub{}\mathclose]$. At each iteration, an intermediate
value $\pivot{}\in\mathopen(\lb{},\ub{}\mathclose]$ is chosen, and the solver
checks if there exists a model with \obj{} lower than \pivot{}. If so, \pivot{}
becomes the new upper bound, otherwise, it becomes the new lower bound. The
search terminates when \lb{} and \ub{} are equal, and the last model found is
optimum. (In continuous domains, e.g., \omlarat{}, to guarantee termination, it
is necessary to interleave binary-search steps with a linear-search
step~\cite{sebastianiOptimizationSMTLAQ2012}).
%
In this paper, we focus on the linear-search strategy, but the analysis applies
to the binary-search strategy as well.

If $\vi$ is built on a combination of theories, then the \T-minimizer is
invoked on $\mu_\T\cup\mu_{ed}$ ---i.e., the subset of $\mu$ containing only
the atoms in \T{} and the interface
(dis-)equalities~\cite{sebastianiOptimizationSMTLAQ2012,sebastianiOptimizationModuloTheories2015}
The implementation of \T-minimizers for \larat{} and \laratint{} is briefly
described in the next paragraph.


\paragraph{T-minimizers.}%
\label{sec:bg:t-minimizers}
A \larat-minimizer{}~\cite{sebastianiOptimizationSMTLAQ2012,sebastianiOptimizationModuloTheories2015} can be implemented
as a simple extension of the Simplex-based
\larat-solver~\cite{dutertreFastLinearArithmeticSolver2006}.
For \laratint{}, a minimizer can be built on top of a branch-and-bound \laratint-solver~\cite{bjornerNZMaximalSatisfaction2014,bjornerNZOptimizingSMT2015,sebastianiPushingEnvelopeOptimization2015}, by replacing the \larat-solver with a \larat-solver\&minimizer to solve each relaxed subproblem.
To find an optimum model within the truth assignment, once a \laratint-model $\M$ is found, a constraint $(\obj < \M(\obj))$ is pushed onto the \laratint-solver, and the search is iteratively refined until no better model exists. An alternative strategy, implemented, e.g., in \optimathsat{}, is the so-called \emph{truncated} optimization~\cite{sebastianiPushingEnvelopeOptimization2015}, where the \laratint-minimizer stops after finding the first model $\M$. Although this model may be sub-optimal for the current truth assignment ---allowing the same assignment to be found again by the \dpll(\T) procedure--- this approach is typically much faster and remains effective in practice.


\begin{remark}
  Importantly, the lazy OMT solving approach allows for an \emph{anytime} behavior, i.e., we
  can interrupt the search at any time and return the best model found so far.
\end{remark}

\subsection{SAT and SMT Enumeration}%
\label{sec:bg:partial-truth-assignments}

SAT enumeration (AllSAT) is the problem of finding all the truth assignments
that propositionally satisfy a propositional formula. \smt{} enumeration
(AllSMT) is the problem of finding all \T-satisfiable truth assignments that
propositionally satisfy a \T-formula. Since a partial assignment can be
extended to $2^k$ total truth assignments, $k$ being the number of unassigned
atoms, finding short partial truth assignments is a key point in reducing both
the number of enumerated truth assignments and the computational time by up to
an exponential factor.

\begin{algorithm}[t]
  \begin{algorithmic}[1]
    \caption[A]{{\minimizeassignment}($\vi, \eta$)\\
      \hspace*{\algorithmicindent}\textbf{Input}:
      CNF formula $\vi$, \T-satisfiable total truth assignment $\eta$ satisfying $\vi$\\
      \hspace*{\algorithmicindent}\textbf{Output}: Reduced (minimal) partial truth assignment $\mu\subseteq\eta$ satisfying $\vi$}%
    \label{alg:minimize}
    \STATE $\mu \leftarrow \eta$
    \FOR{$\ell\in\mu$}
    \IF{$\mu\setminus\set{\ell}$ satisfies all clauses in $\vi$}
    \STATE $\mu \leftarrow \mu \setminus \set{\ell}$
    \ENDIF
    \ENDFOR
    \RETURN $\mu$
  \end{algorithmic}
\end{algorithm}

Many enumeration algorithms find total truth assignments, and then extract
partial truth assignments from them by some reduction procedure. A basic
reduction procedure is illustrated in~\Cref{alg:minimize}. It consists in
iteratively dropping literals one-by-one from the truth assignment, checking if
it still satisfies the formula. The resulting partial assignment is minimal,
i.e., it cannot be further reduced without violating the satisfaction of the
formula. Notice that the order in which literals are dropped can have a
significant impact on the effectiveness of the reduction procedure.
\section{An Analysis of Enumeration in OMT}%
\label{sec:analysis}

In the following, to simplify the notation and the presentation, we refer to
one single theory \T{}, but the results can be straightforwardly extended to
combinations of theories.

As described in~\sref{sec:omt-solving}, a basic \omt{} solving schema involves
the interaction of a combinatorial and a theory-specific optimization
components. In the combinatorial component, an \smt{} solver enumerates
\T{}-satisfiable truth assignments that propositionally satisfy the problem
formula $\vi$ conjoined with increasingly tighter bounds on the cost of the
optimum solution. In the theory-specific component, a \T-minimizer finds a
\T{}-model of minimum cost within the constraints imposed by the given truth
assignment. This model is then used to tighten the upper bound for the cost of
the optimum model and continue the search, until the formula is found
unsatisfiable.

Since the enumeration is based on the CDCL(\T)
schema~\cite{marques-silvaConflictDrivenClauseLearning2021}, these truth
assignments are typically \emph{total}, i.e., they assign a truth value to each
atom of the formula. 
%
%
However, we point out that total truth assignments can often over-constrain the
search space for the optimum model, whereas relying on \emph{partial} truth
assignments can be much more effective. Intuitively, \emph{by removing from the
    current satisfying truth assignment \T{}-constraints that are not strictly
    necessary for the propositional satisfaction of the formula, we enlarge the
    area within which the optimum model is searched, thus increasing the chances of
    finding a better optimum model.}
This means that the solver can add a tighter upper bound to the cost of the
global optimum, potentially reducing the number of search iterations needed to
find it, and consequently the overall solving time. Moreover, this improvement
can be crucial for anytime OMT solving, as it allows the solver to converge
faster to better solutions within the given time limit.

We illustrate this idea in the following example.


\begin{figure}[t]
    \input{figures/partial-assignments/preamble.tex}
    \centering
    \begin{subfigure}{0.33\textwidth}%
        \resizebox{\columnwidth}{!}{\begin{tikzpicture}
    \path (0,0) pic {planes};
    \fill[plane, opacity=\planeopacity] (BL) -- (TL) -- (TR) -- cycle;
    \fill[plane, opacity=\planeopacity] (13/3,\mY) -- (2, \MY) -- (TL) -- (BL) -- cycle;
    \fill[plane, opacity=\planeopacity] (\mX,2) -- (\MX, 2) -- (BR) -- (BL) -- cycle;
    \fill[plane, opacity=\planeopacity] (-2,\mY) -- (BR) -- (TR) -- (-2,\MY) -- cycle;
    \fill[plane, opacity=\planeopacity] (4,\MY) -- (4, \mY) -- (BL) -- (TL) -- cycle;
    \draw[dashed, name path=plane1] (BL) -- (TR) node[above left] {\LARGE$2x-3y\leq 6$};
    \draw[dashed, name path=plane2] ( 13/3,\mY) -- (2, \MY) node[above] {\LARGE$y \leq -3x + 9$};
    \draw[dashed, name path=plane3] (-3, 2) -- (23/3, 2) node[below left] {\LARGE$y \leq 2$};
    \draw[dashed, name path=plane4] (-2,3) -- (-2,-4) node[below] {\LARGE$\neg(x < -2)$};
    \draw[dashed, name path=plane5] (4,\MY) -- ( 4,\mY) node[below] {\LARGE$x \leq 4$};
    \path[name intersections={of=plane3 and plane4, by={A}}];
    \path[name intersections={of=plane2 and plane3, by={B}}];
    \path[name intersections={of=plane5 and plane3, by={C}}];
    \path[name intersections={of=plane1 and plane3, by={D}}];
    \path[name intersections={of=plane1 and plane5, by={E}}];
    \path[name intersections={of=plane1 and plane2, by={F}}];
    \path[name intersections={of=plane1 and plane4, by={G}}];
    \draw[very thick] (A) -- (B) -- (F) -- (G) -- cycle;
    \fill[opacity=.4, top color=white, bottom color=blue, shading angle=64.76] (A) -- (B) -- (F) -- (G) -- cycle;
    \path (0,0) pic {axes};
    \draw[fill=red]   (F) circle (1.5mm);
\end{tikzpicture}}%
        \caption{Total assignment $\mu$~\eqref{eq:omt-partial-assignments:total-assignment:mu}}%
        \label{fig:omt-partial-assignments:step1}%
    \end{subfigure}%
    \begin{subfigure}{0.33\textwidth}%
        \resizebox{\columnwidth}{!}{\begin{tikzpicture}
    \path (0,0) pic {planes};
    \fill[plane, opacity=\planeopacity] (BL) -- (TL) -- (TR) -- cycle;
    \fill[plane, opacity=\planeopacity] (\mX,2) -- (\MX, 2) -- (BR) -- (BL) -- cycle;
    \fill[plane, opacity=\planeopacity] (-2,\mY) -- (BR) -- (TR) -- (-2,\MY) -- cycle;
    \fill[plane, opacity=\planeopacity] (4,\MY) -- (4, \mY) -- (BL) -- (TL) -- cycle;
    \draw[dashed, name path=plane1] (BL) -- (TR) node[above left] {\LARGE$2x-3y\leq 6$};
    \draw[dashed, name path=plane2, opacity=0.3] ( 13/3,\mY) -- (2, \MY) node[above] {\LARGE$y \leq -3x + 9$};
    \draw[dashed, name path=plane3] (-3, 2) -- (23/3, 2) node[below left] {\LARGE$y \leq 2$};
    \draw[dashed, name path=plane4] (-2,3) -- (-2,-4) node[below] {\LARGE$\neg(x < -2)$};
    \draw[dashed, name path=plane5] (4,\MY) -- ( 4,\mY) node[below] {\LARGE$x \leq 4$};
    \path[name intersections={of=plane3 and plane4, by={A}}];
    \path[name intersections={of=plane2 and plane3, by={B}}];
    \path[name intersections={of=plane5 and plane3, by={C}}];
    \path[name intersections={of=plane1 and plane3, by={D}}];
    \path[name intersections={of=plane1 and plane5, by={E}}];
    \path[name intersections={of=plane1 and plane2, by={F}}];
    \path[name intersections={of=plane1 and plane4, by={G}}];
    \draw[very thick] (A) -- (C) -- (E) -- (G) -- cycle;
    \fill[opacity=.4, top color=white, bottom color=blue, shading angle=64.76] (A) -- (C) -- (E) -- (G) -- cycle;
    \path (0,0) pic {axes};
    \draw[fill=red]   (E) circle (1.5mm);
\end{tikzpicture}}%
        \caption{Partial assignment $\muprime$~\eqref{eq:omt-partial-assignments:total-assignment:muprime}}%
        \label{fig:omt-partial-assignments:step2}%
    \end{subfigure}%
    \begin{subfigure}{0.33\textwidth}%
        \resizebox{\columnwidth}{!}{\begin{tikzpicture}
    \path (0,0) pic {planes};
    \fill[plane, opacity=\planeopacity] (BL) -- (TL) -- (TR) -- cycle;
    \fill[plane, opacity=\planeopacity] (\mX,2) -- (\MX, 2) -- (BR) -- (BL) -- cycle;
    \fill[plane, opacity=\planeopacity] (-2,\mY) -- (BR) -- (TR) -- (-2,\MY) -- cycle;
    \draw[dashed, name path=plane1] (BL) -- (TR) node[above left] {\LARGE$2x-3y\leq 6$};
    \draw[dashed, name path=plane2, opacity=0.3] ( 13/3,\mY) -- (2, \MY) node[above] {\LARGE$y \leq -3x + 9$};
    \draw[dashed, name path=plane3] (-3, 2) -- (23/3, 2) node[below left] {\LARGE$y \leq 2$};
    \draw[dashed, name path=plane4] (-2,3) -- (-2,-4) node[below] {\LARGE$\neg(x < -2)$};
    \draw[dashed, name path=plane5, opacity=0.3] (4,\MY) -- ( 4,\mY) node[below] {\LARGE$x \leq 4$};
    \path[name intersections={of=plane3 and plane4, by={A}}];
    \path[name intersections={of=plane2 and plane3, by={B}}];
    \path[name intersections={of=plane5 and plane3, by={C}}];
    \path[name intersections={of=plane1 and plane3, by={D}}];
    \path[name intersections={of=plane1 and plane5, by={E}}];
    \path[name intersections={of=plane1 and plane2, by={F}}];
    \path[name intersections={of=plane1 and plane4, by={G}}];
    \draw[very thick] (A) -- (C) -- (D) -- (G) -- cycle;
    \fill[opacity=.4, top color=white, bottom color=blue, shading angle=64.76] (A) -- (D) -- (G) -- cycle;
    \path (0,0) pic {axes};
    \draw[fill=red]   (D) circle (1.5mm);
\end{tikzpicture}}
        \caption{Partial assignment $\mupprime$~\eqref{eq:omt-partial-assignments:total-assignment:mupprime}}%
        \label{fig:omt-partial-assignments:step3}%
    \end{subfigure}%
    \caption{
        Graphical representation of~\Cref{ex:omt-partial-assignments}. For each step, the half-planes representing the constraints in the truth assignment are delimited by dashed lines and colored in grey. The intersection of these constraints is colored in blue, with a gradient that follows the value of $\obj$ (the lower the value of $\obj$, the more intense the color), and the red dot represents the optimum model found within this region.
    }%
    \label{fig:omt-partial-assignments}%
\end{figure}

\begin{example}%
    \label{ex:omt-partial-assignments}
    Consider the \omlarat{} problem \pair{\vi}{\obj} where \vi{} is the formula in~\eqref{eq:smt} in~\Cref{ex:smt}, and $\obj\defas -2x$.
    Consider the following scenario, which is graphically represented in~\Cref{fig:omt-partial-assignments}. Consider the \larat{}-satisfiable total truth assignment that propositionally satisfies $\vi$:
    \begin{equation}%
        \label{eq:omt-partial-assignments:total-assignment:mu}
        \mu\defas\set{(2x-3y\leq 6),(y\leq 2),\neg(x<-2),(y\leq-3x+9),(x\leq 4)}.
    \end{equation}
    The optimum model of $\mu$ is \set{x\mapsto{}3,y\mapsto{}0} with $\obj=-6$ (\Cref{fig:omt-partial-assignments:step1}). 
    We notice that some literals in $\mu$ are not strictly necessary for
    propositionally satisfying $\vi$. In fact, we only need one true literal in
    each clause to propositionally satisfying the formula. Not every drop is
    equally effective, though. For instance, if we drop $(x\leq 4)$, $(y\leq 2)$,
    or $\neg(x < 2)$, then we get a truth assignment with the same optimum as
    $\mu$, since these literals don't ``oppose'' to the optimization of $\obj$ in
    $\mu$. Dropping at least one of the other two literals, instead, leads to a
    truth assignment with a better optimum model. For instance, if we drop
    $(y\leq-3x+9)$, we get:
    \begin{equation}%
        \label{eq:omt-partial-assignments:total-assignment:muprime}
        \muprime\defas\mu\setminus\set{(y\leq-3x+9)}=\set{(2x-3y\leq 6),(y\leq 2),\neg(x<-2),(x\leq 4)}.
    \end{equation}
    The optimum model of $\muprime$ is \set{x\mapsto{}4,y\mapsto{}2/3} with
    $\obj=-8$ (\Cref{fig:omt-partial-assignments:step2}).
    At this point, two constraints, $(x\leq 4)$ and $(2x-3y\leq 6)$, oppose to the
    optimization of $\obj$, and either of them can be safely dropped. Assume that
    we drop $(x \leq 4)$, then we get:
    \begin{equation}%
        \label{eq:omt-partial-assignments:total-assignment:mupprime}
        \mupprime\defas\muprime\setminus\set{(x \leq
            4)}=\set{(2x-3y\leq 6),(y\leq 2),\neg(x<-2)}
    \end{equation}
    with optimum model \set{x\mapsto{}6,y\mapsto{}2} and $\obj=-12$ (\Cref{fig:omt-partial-assignments:step3}). Now, the only literals opposing to the optimization of $\obj$ are $(2x-3y\leq 6)$ and $(y\leq 2)$, but none of them can be dropped. Hence, no further improvement can be obtained.
\end{example}

In general, partial truth assignments have an optimum model that is necessarily
better or equal to that of the total truth assignments extending them. Since
multiple partial truth assignments can be obtained from a total one, the choice
of which constraints to drop can be crucial to improve the quality of the
optimum model found.
\section{Exploiting Partial Truth Assignments in OMT}%
\label{sec:approach}
\begin{algorithm}[t]
    \newcommand{\res}{\textsf{res}}
    \begin{algorithmic}[1]
        \caption[A]{{\sc Linear-search OMT with partial assignments}($\vi, \obj$)\\
            \hspace*{\algorithmicindent}\textbf{Input}:
            Formula $\vi$, objective $\obj$\\
            \hspace*{\algorithmicindent}\textbf{Output}: $\satres/\unsatres$, optimum model $\M$}%
        \label{alg:omt-partial}
        \STATE \makebox[.5cm][c]{$\M$}$\gets \emptyset$ \algorithmiccomment{Best model found so far}
        \STATE \makebox[.5cm][c]{$\ub$}$\gets \infty$ \algorithmiccomment{Current upper bound}
        \STATE \makebox[.5cm][c]{$\res$}$\gets \satres$ \algorithmiccomment{Status of the search}
        \WHILE{$\res = \satres$}
        \STATE $\tuple{\res,\eta} \gets \incrementalsmt(\vi\wedge(\obj<\ub))$
        \IF{$\res = \satres$}
        \STATE\label{alg:line:omt-partial:minimize} \makebox[.5cm][c]{$\color{blue}\mu$}$\color{blue}\gets\omtminimizeassignment(\vi,\eta,\obj)$
        \STATE\label{alg:line:omt-partial:tminimize} \makebox[.5cm][c]{$\M$}$\gets \minimize(\mu,\obj)$
        \STATE \makebox[.5cm][c]{$\ub$}$\gets \M(\obj)$
        \ENDIF
        \ENDWHILE
        \IF{$\M = \emptyset$}
        \RETURN $\tuple{\unsatres,\emptyset}$
        \ELSE
        \RETURN $\tuple{\satres,\M}$
        \ENDIF
    \end{algorithmic}
\end{algorithm}

The general schema of our approach is presented in~\Cref{alg:omt-partial}. This
algorithm is a variant of the basic \omt{} linear-search
schema~\cite{sebastianiOptimizationSMTLAQ2012,sebastianiOptimizationModuloTheories2015}
described in~\sref{sec:bg:omt}. The main difference is the call to the
$\omtminimizeassignment$ procedure (line~\ref{alg:line:omt-partial:minimize}),
which is responsible for reducing the truth assignment to be fed to the
\T-minimizer, provided that the resulting partial truth assignment still
propositionally satisfies the formula. Depending on the implementation of this
procedure, the assignment-reduction strategy can be more or less effective in
improving the search for the global optimum.

In \sref{sec:approach:basic-assignment-minimization} and
\sref{sec:approach:guided-assignment-minimization}, we describe two possible
implementations of this procedure.


\subsection{Basic Assignment Reduction}%
\label{sec:approach:basic-assignment-minimization}

The first approach is to reduce the truth assignment using~\Cref{alg:minimize}
in~\sref{sec:bg:partial-truth-assignments}, i.e., iterating over all the
literals in the current truth assignment $\eta$, and dropping them one by one,
if possible.
A straightforward improvement is to only try to drop \T-literals, since they
are the ones that, if dropped, can potentially enlarge the area within which
the optimum \T-model is searched. In the case of theory combination, we drop
only literals from the theory \T{} of the ``$\le$'' symbol w.r.t.\ which we
minimize. Possibly, heuristics can be used to choose an appropriate dropping
order; in our implementation, we used the default strategy that follows the
appearance order of the atoms in the formula. This procedure is simple and
general, and comes with a limited overhead, as each truth assignment is scanned
only once to find the literals to drop, and the \T-minimizer is called only
once for each candidate assignment.

This approach, however, might not be very effective in practice, as it
``blindly'' removes literals from the truth assignment without taking into
account the properties of the \omt{} search strategy. In particular, it may
drop literals that are not relevant for the optimization, enlarging the search
space in a direction that does not help improve the objective, possibly
preventing the removal of more relevant literals.

\subsection{Guided Assignment Reduction}%
\label{sec:approach:guided-assignment-minimization}
We propose an ad-hoc assignment-reduction technique for \omt{} solving, which
is outlined in~\Cref{alg:omt-minimize-assignment-guided}.
%
%
%
Suppose that, after the \T-minimizer has found a minimum model within the
current truth assignment $\mu$
(line~\ref{alg:omt-minimize-assignment-guided:line:minimize1}), it returns also
one (or more) literal(s) that limit the current minimum
(line~\ref{alg:omt-minimize-assignment-guided:line:propose1}). These literals
are part of some (possibly minimal) $\muprime\subseteq\mu$ such that
$\muprime\cup\set{\obj<\M(\obj)}$ is \T-unsatisfiable. Intuitively, the removal
of any literal $\ell\in\muprime$ is very likely to lead to a better optimum
model, provided that $\mu\setminus\set{\ell}$ still propositionally satisfies
$\vi$ (line~\ref{alg:omt-minimize-assignment-guided:line:ifcandrop}).

We can then iteratively drop these literals and re-run the \T-minimizer, until
no more literals can be dropped
(lines~\ref{alg:omt-minimize-assignment-guided:line:while}--\ref{alg:omt-minimize-assignment-guided:line:propose2}).
Notice that instead of \minimize{}, here we call \minimizestar{}, suggesting
that also a relaxed optimization algorithm could be used. In the next
paragraph, we describe how this can be exploited for \laratint.
\begin{algorithm}[t]
    \begin{algorithmic}[1]
        \caption[A]{\omtminimizeassignmentguided($\vi, \eta$, $\obj$)\\
            \hspace*{\algorithmicindent}\textbf{Input}:
            Formula $\vi$, \T-satisfiable total truth assignment $\eta$ satisfying $\vi$, objective $\obj$\\
            \hspace*{\algorithmicindent}\textbf{Output}: Reduced truth assignment $\mu\subseteq\eta$ satisfying $\vi$}%
        \label{alg:omt-minimize-assignment-guided}
        \STATE\makebox[.5cm][c]{$\mu$}$\gets\eta$
        \STATE\makebox[.5cm][c]{$\M$}$\gets\minimizestar(\mu,\obj)$\label{alg:omt-minimize-assignment-guided:line:minimize1}
        \STATE\makebox[.5cm][c]{$\ell$}$\gets\proposelit()$\label{alg:omt-minimize-assignment-guided:line:propose1}
        \WHILE{$\ell\neq\bot$}\label{alg:omt-minimize-assignment-guided:line:while}
        \IF{$\mu\setminus\set{\ell}$ satisfies all clauses in $\vi$}\label{alg:omt-minimize-assignment-guided:line:ifcandrop}
        \STATE \makebox[.5cm][c]{$\mu$}$\gets \mu \setminus \set{\ell}$\label{alg:omt-minimize-assignment-guided:line:drop}
        \STATE \makebox[.5cm][c]{$\M$}$\gets\minimizestar(\mu,\obj)$\label{alg:omt-minimize-assignment-guided:line:minimize2}
        \ENDIF
        \STATE $\ell\gets\proposelit()$\label{alg:omt-minimize-assignment-guided:line:propose2}
        \ENDWHILE
        \RETURN $\mu$
    \end{algorithmic}
\end{algorithm}

Notice that, in \Cref{alg:omt-minimize-assignment-guided}, we only drop one
literal before every optimization call; nevertheless, in principle, more
literals could be dropped. Experimentally, we found that dropping more than one
literal was not beneficial. The reason is that, generally, when a literal is
removed, most of the other literals in $\muprime$ do not limit the current
minimum anymore. Hence, their removal not only does not lead to a better
optimum model, but also can prevent the removal of other more-relevant
literals.

\paragraph{On proposing literals to drop.}
For an arbitrary theory \T, a generic implementation for \proposelit{}
(\Cref{alg:omt-minimize-assignment-guided},
lines~\ref{alg:omt-minimize-assignment-guided:line:propose1},~\ref{alg:omt-minimize-assignment-guided:line:propose2})
could be as follows. Once a minimum model $\M$ for $\mu$ is found, invoke the
\T-solver on $\mu \wedge (\obj < \M(\obj))$ and return a (possibly-minimal)
conflict set. For some theories, ad-hoc (possibly heuristic) procedures can be
employed. We describe two such techniques for \larat{} and \laratint{}.

%
As we recalled in~\sref{sec:bg:omt}, a \larat{}-minimizer can be implemented as
a variant of the Simplex
method~\cite{dutertreFastLinearArithmeticSolver2006,sebastianiOptimizationSMTLAQ2012},
by which an optimum model is always found on a vertex of the polytope defined
by the conjunction of \larat-constraints on which it is invoked. Thus, in this
case, the candidate constraints to be dropped are those that form such a
vertex. This information can be easily obtained from the Simplex
tableau~\cite{dutertreFastLinearArithmeticSolver2006}.

For \laratint{}, directly using a \laratint-minimizer{} and then identifying
the limiting constraints presents some challenges:
\begin{enumerate*}[label=(\alph*)]
    \item a single call to the branch-and-bound-based \laratint-minimizer is worst-case
          exponential, hence calling it multiple times (as
          in~\Cref{alg:omt-minimize-assignment-guided}) is not feasible;
    \item extracting the limiting constraints from a \laratint-minimizer is not
          straightforward.
\end{enumerate*}
We thus propose proceeding as follows. First, for the procedure \minimizestar{} a \larat-minimizer{} is invoked on the relaxation of $\mu$.
Then, the procedure \proposelit{} can be implemented as for \larat{}.
The intuition here is that reasoning on the relaxation of $\mu$ is much easier and cheaper ---especially if incremental calls are used--- and still allows for enlarging the search area in a favourable direction for the \laratint-minimizer. We remark that, after the assignment reduction, in \Cref{alg:omt-partial} (line~\ref{alg:line:omt-partial:tminimize}), the complete \laratint-minimizer is called.

As a last aspect, we suggest that, if the \T-minimizer is able to find an
optimum model $\mu$, then we can use these limiting literals $l_1,\dots,l_n$
also to learn a theory lemma $(\neg(\obj < \ub) \vee \neg{}l_1 \vee \ldots \vee
    \neg{}l_n)$ that blocks truth assignments that we know are not better than the
current upper bound $\ub$ ---thus preventing useless calls to the \T-solver.
The idea is that, in order to find a model with $\obj < \ub$, we need to assign
at least one of these literals to false. This is the case of \larat{}, but not
of \laratint{} if the truncated minimization method is used
(see~\sref{sec:bg:t-minimizers}).
\section{Experimental evaluation}%
\label{sec:experiments}

We implemented the above algorithms in the OMT solver
\optimathsat{}~\cite{sebastianiOptiMathSATToolOptimization2020}, which is built
on top of the \mathsatfive{} SMT solver~\cite{mathsat5_tacas13}.
We evaluated the proposed strategies on a set of \omlarat{}, \omlaratint, and
\omlaratplusa{} benchmarks coming from different sources, evaluating both
solving time for optimum solving, and the quality of the solutions found within
the given timeout for anytime solving.
All the experiments were run on an Intel Xeon Gold 6238R @ 2.20GHz 28 Core
machine with 128 GB of RAM, running Ubuntu Linux 22.04. The timeout was set at
1200s. The tool, benchmarks and results are available at
\url{https://optimathsat.disi.unitn.it/resources/optimathsat-partial-assignments.tar.gz}.

\subsection{Benchmarks}%
\label{sec:experiments:benchmarks}
We evaluated the proposed strategies on two classes of \omlarat{} benchmarks:
OMT-encoded optimal temporal
planning~\cite{panjkovicExpressiveOptimalTemporal2023,panjkovicAbstractActionScheduling2024}
and strip-packing
problems~\cite{sebastianiOptimizationSMTLAQ2012,sebastianiOptimizationModuloTheories2015}.
We also modified the strip-packing benchmarks to use \omlaratint{} and
\omlaratplusa{} encodings, to evaluate the effectiveness of our strategies in
these theories.

\paragraph{Optimal Temporal Planning.}
In~\cite{panjkovicExpressiveOptimalTemporal2023,panjkovicAbstractActionScheduling2024},
the authors proposed a way to encode optimal temporal planning problems into a
sequence of \omlarat{} problems. Each problem encodes a bounded version of the
problem up to a fixed horizon, with additional abstract actions representing an
over-approximation of the plans beyond the bound, minimizing the makespan,
i.e., the total time taken to reach the goal. If the optimal plan is found
without using the abstract actions, then the plan is optimum for the original
problem. Otherwise, the horizon is increased, and the process is repeated. We
generated problems using the industrial problems Majsp (80 instances),
MajspSimplified (80 instances), and Painter (30
instances)~\cite{panjkovicAbstractActionScheduling2024}, with increasing
horizon $h\in\set{5, 10, 15, 20, 25, 30, 35, 40}$, for a total of 1520
instances.

\paragraph{Strip-packing.}
The strip-packing problem (SP) requires arranging $N$ rectangles, each with a
specific width $W_i$ and height $H_i$, into a strip of fixed height $H$ and
unlimited length. The goal is to minimize the length $L$ of the used part of
the strip, ensuring that all rectangles are placed without overlap or rotation.
An \omlarat{} encoding for SP was proposed
in~\cite{sebastianiOptimizationModuloTheories2015}.
Following~\cite{sebastianiOptimizationModuloTheories2015}, we sampled $H_i$
uniformly in $(0,1]$, $W_i$ in $(1, 2]$, and set $H=\sqrt{N}/2$.

We also generated \omlaratint{} SP problems, by randomly choosing with equal
probability whether encoding the coordinates of each rectangle with integer or
rational variables.
Finally, we generated \omlaratplusa{} encodings for SP, by simply replacing the
variables $x_i$ in the \omlarat{} encoding with a \laratplusa{} term
$\mathit{read}(x, i + \mathit{offset})$, where $x$ is an array mapping from
rationals to rationals, $i$ is a constant indicating the index of the
rectangle, and $\mathit{offset}$ is a fresh rational variable.

For each of these encodings, we generated 25 random SP problems for each value
of $N\in\set{25,50,75,100}$, for a total of $100$ instances per encoding.

\subsection{Results}%
\label{sec:experiments:results}

\begin{figure}[t]
    \begin{subfigure}{\textwidth}%
        \centering
        \includegraphics[width=.6\columnwidth]{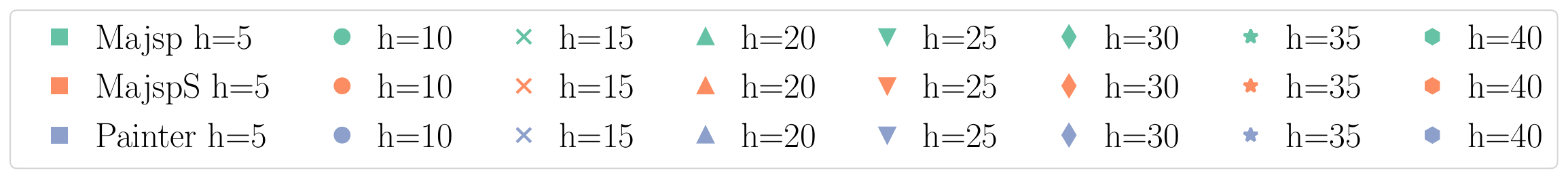}%
    \end{subfigure}
    \begin{subfigure}{\textwidth}%
        \begin{tabularx}{\textwidth}{cc|c}
            \includegraphics[width=.33\columnwidth]{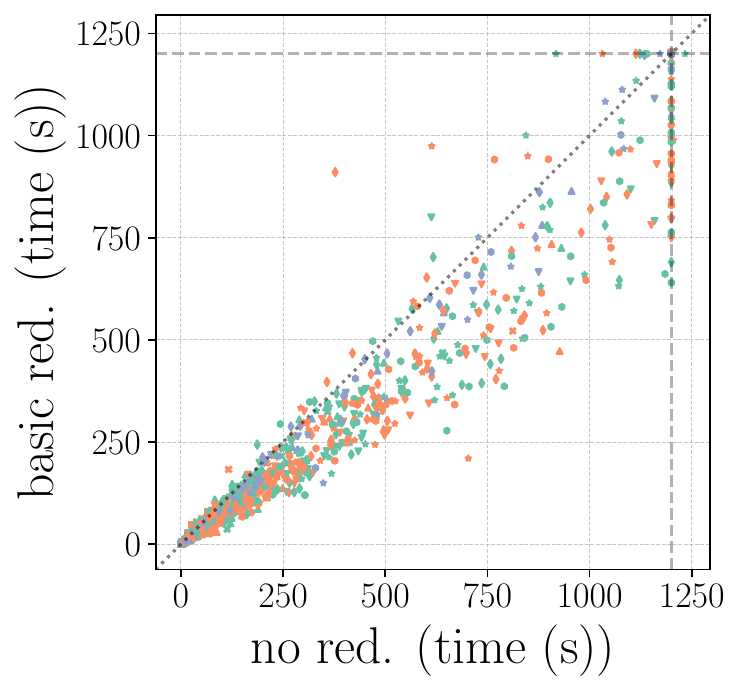}                                   &
            \includegraphics[width=.33\columnwidth]{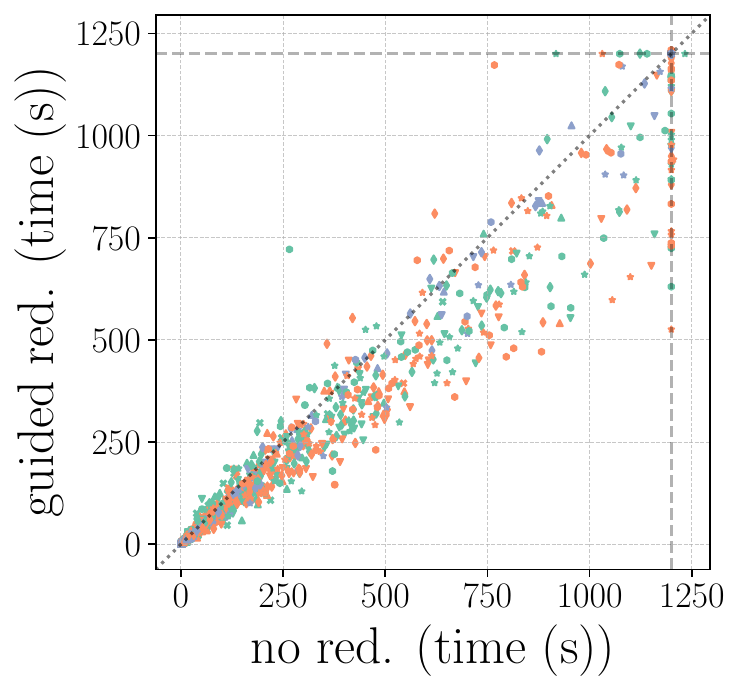}                         &
            \includegraphics[width=.33\columnwidth]{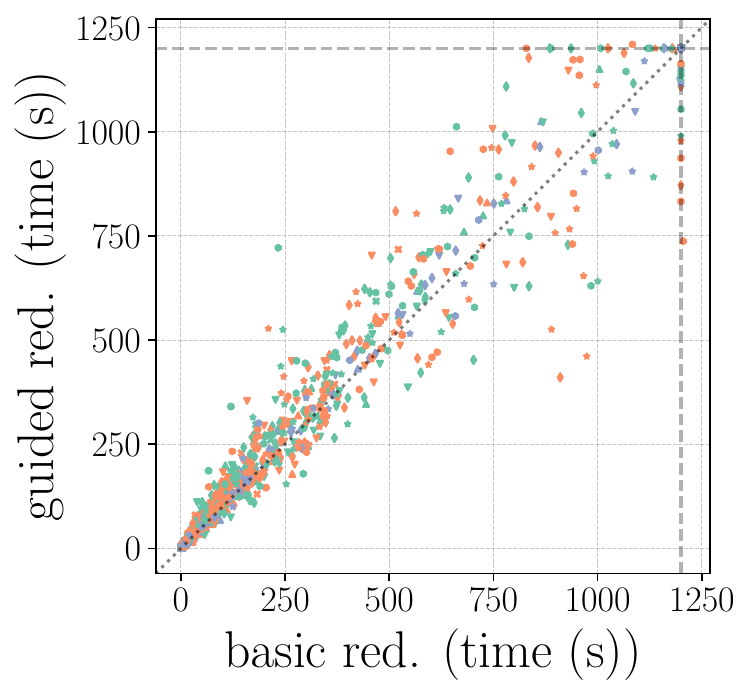}              \\
            \includegraphics[width=.33\columnwidth]{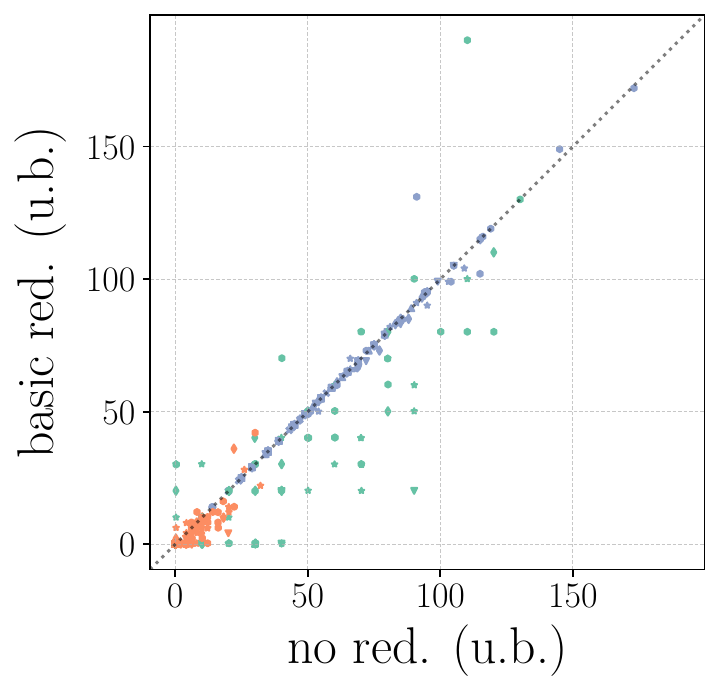}                                 &
            \includegraphics[width=.33\columnwidth]{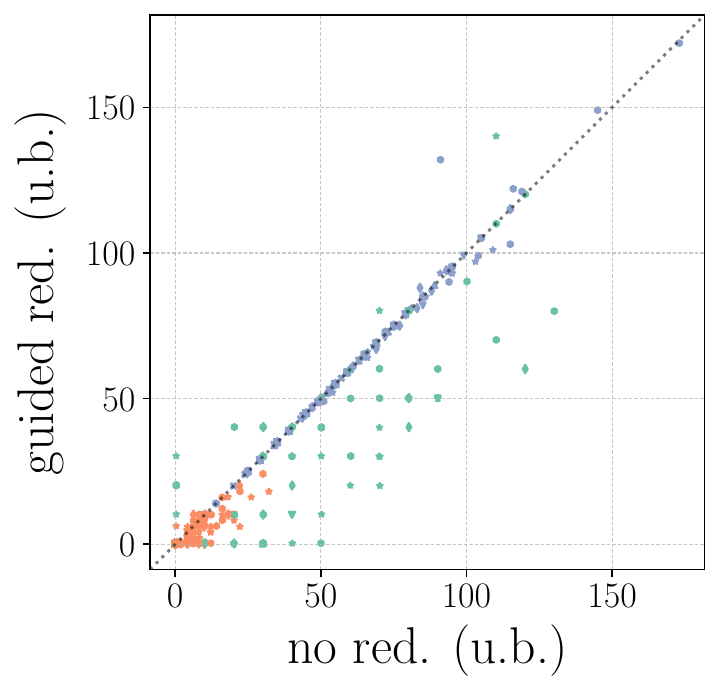}                       &
            \includegraphics[width=.33\columnwidth]{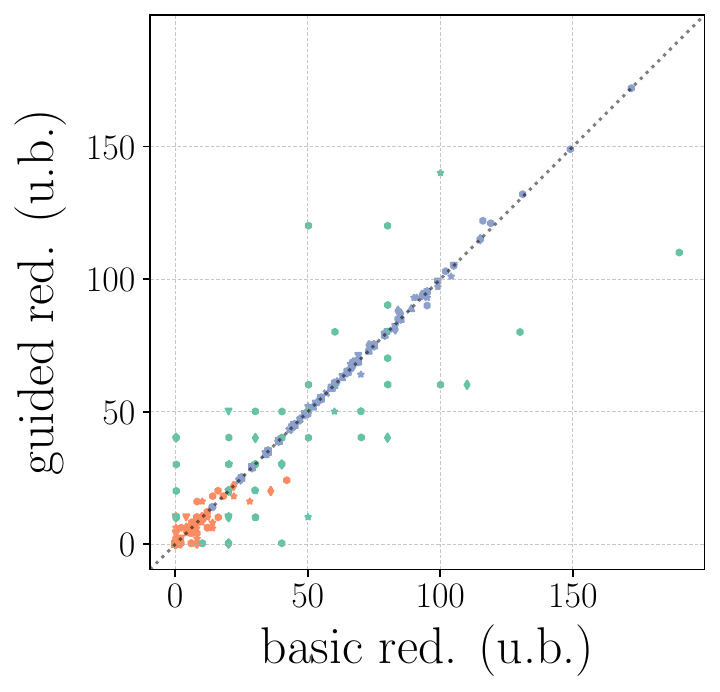}            \\
            \includegraphics[width=.33\columnwidth]{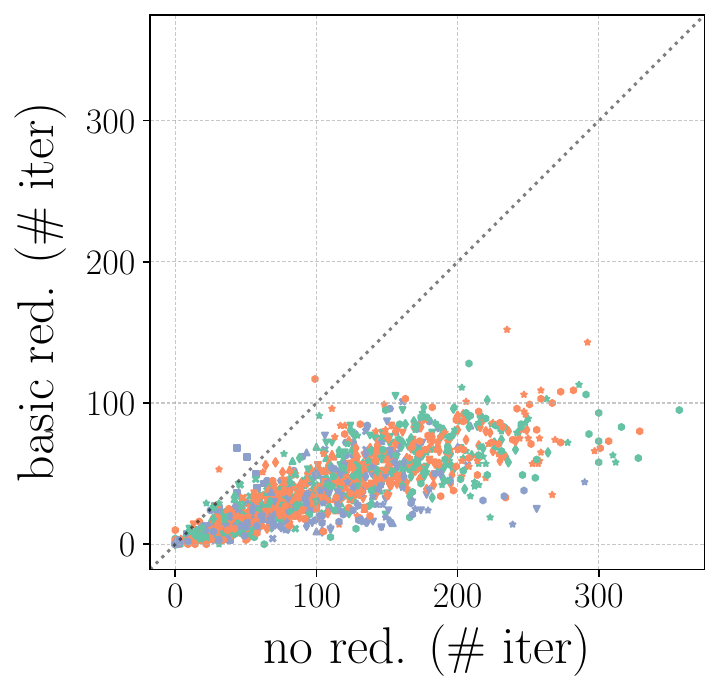}           &
            \includegraphics[width=.33\columnwidth]{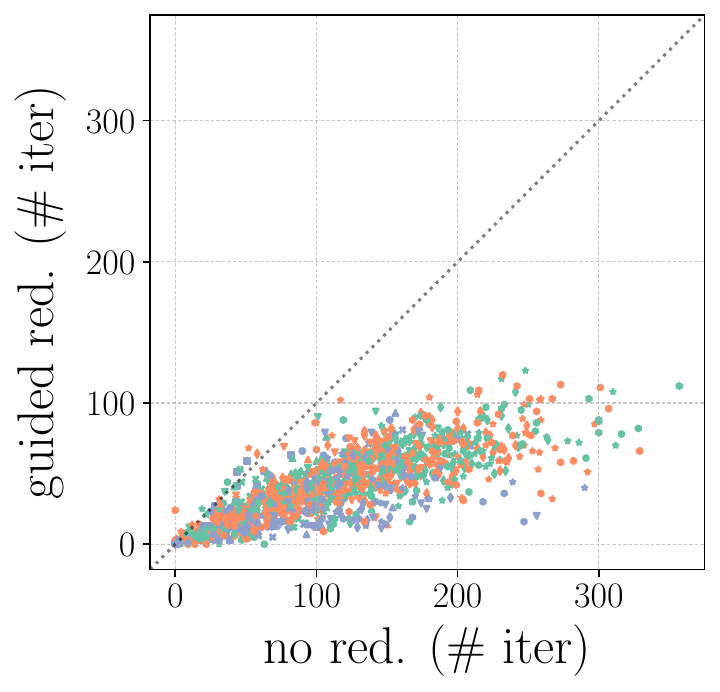} &
            \includegraphics[width=.33\columnwidth]{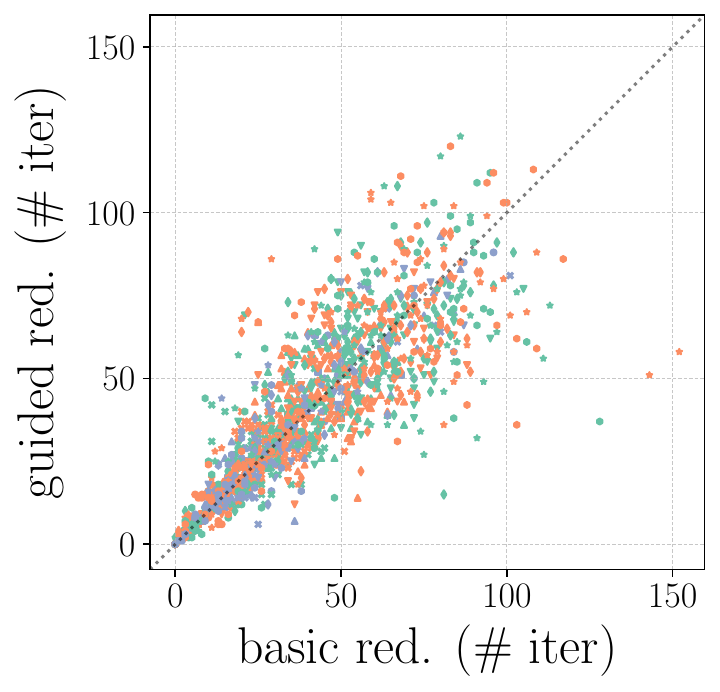}
        \end{tabularx}
    \end{subfigure}

    \caption{
        Results on \omlarat-encoded optimal temporal planning problems.
    }%
    \label{fig:plot:planning}%
\end{figure}

\begin{figure}[t]
    \begin{subfigure}{\textwidth}%
        \centering
        \includegraphics[width=.6\columnwidth]{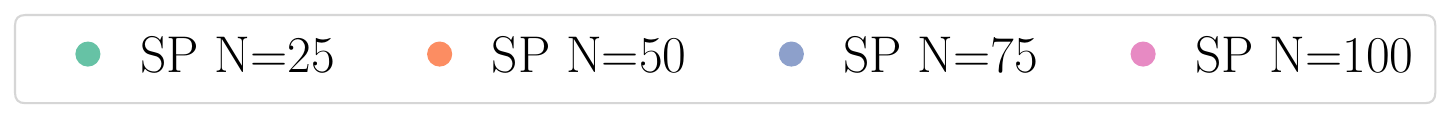}%
    \end{subfigure}
    \begin{subfigure}{\textwidth}%
        \begin{tabularx}{\textwidth}{cc|c}
            \includegraphics[width=.33\columnwidth]{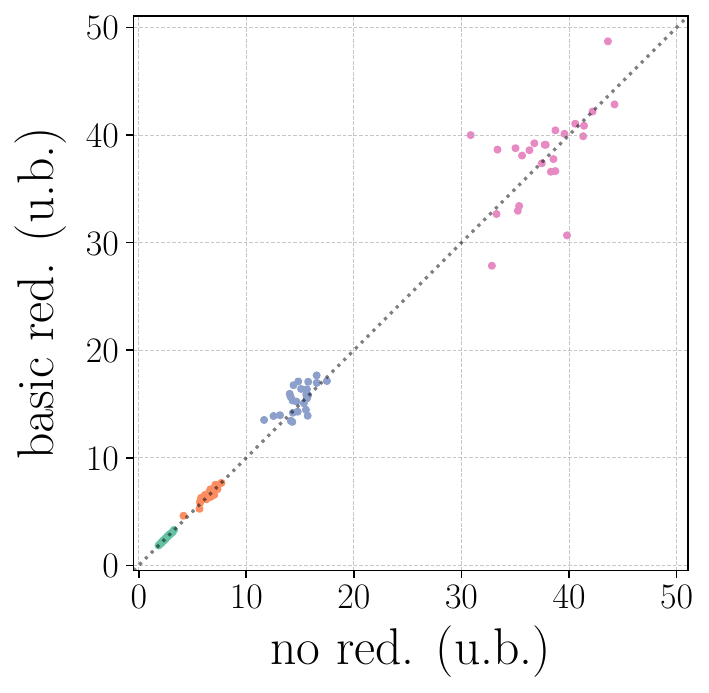}                                 &
            \includegraphics[width=.33\columnwidth]{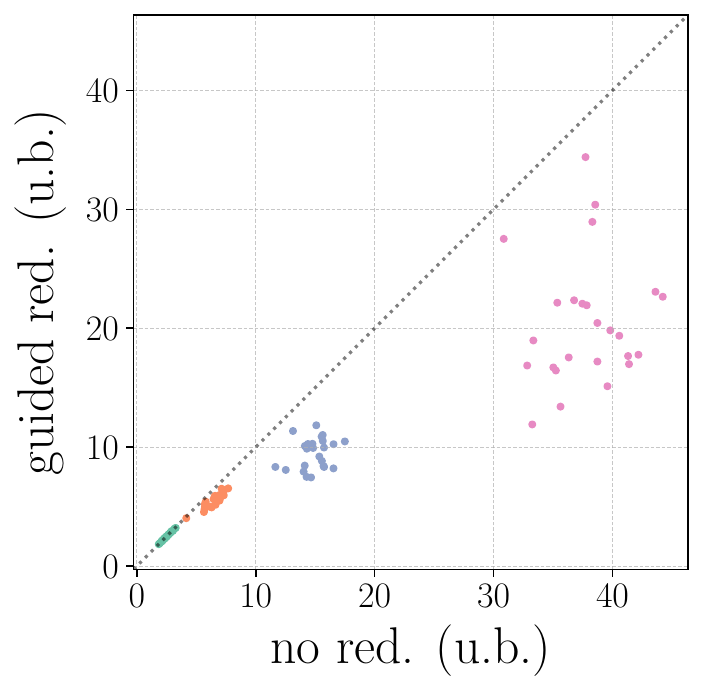}                       &
            \includegraphics[width=.33\columnwidth]{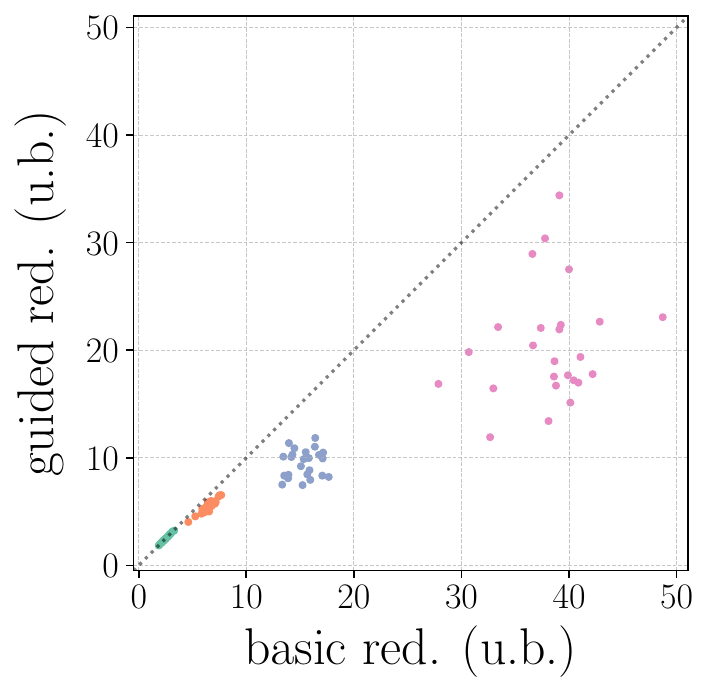}            \\
            \includegraphics[width=.33\columnwidth]{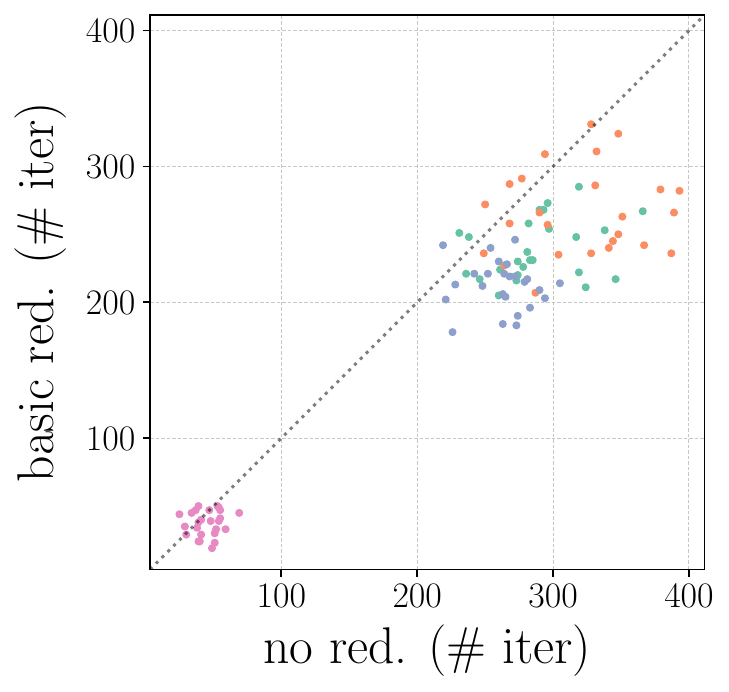}           &
            \includegraphics[width=.33\columnwidth]{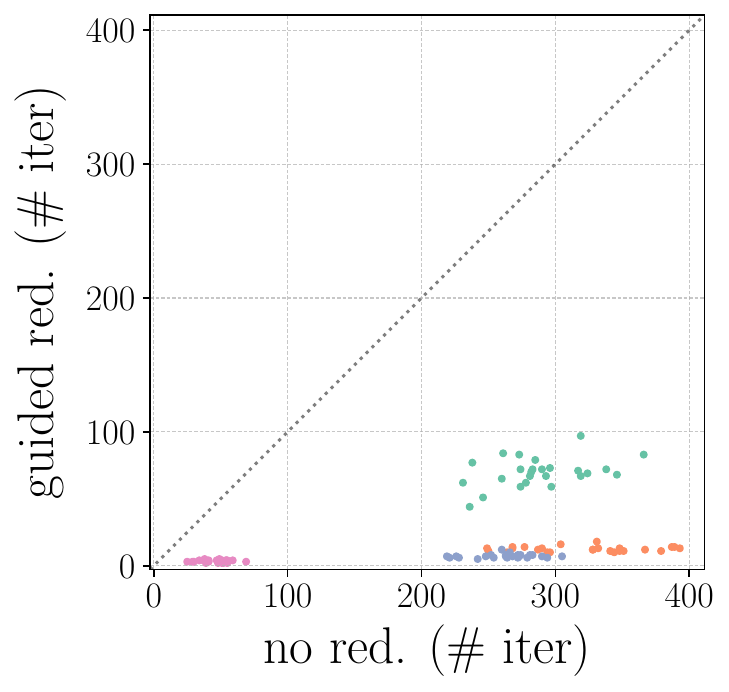} &
            \includegraphics[width=.33\columnwidth]{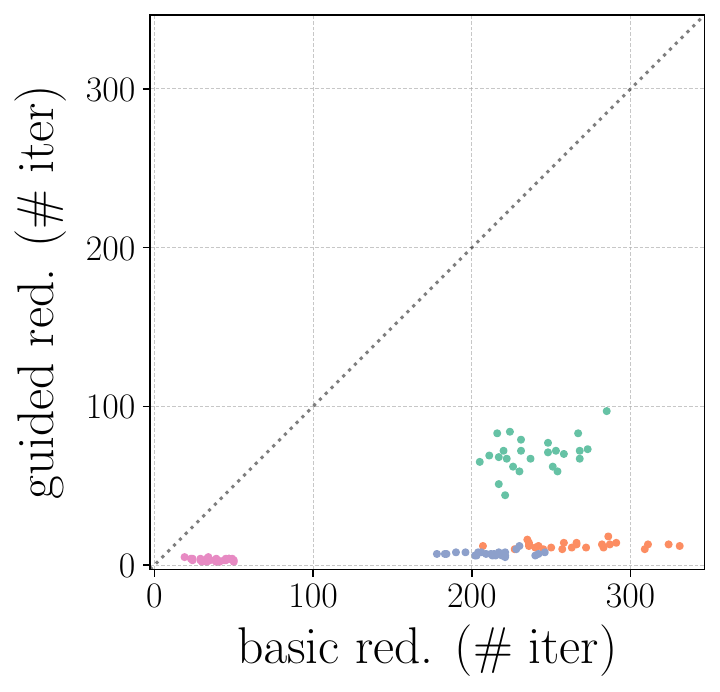}
        \end{tabularx}
    \end{subfigure}
    \caption{Results on \omlarat-encoded strip-packing problems.}%
    \label{fig:plot:sp:lra}%
\end{figure}

\begin{figure}[t]
    \begin{subfigure}{\textwidth}%
        \centering
        \includegraphics[width=.6\columnwidth]{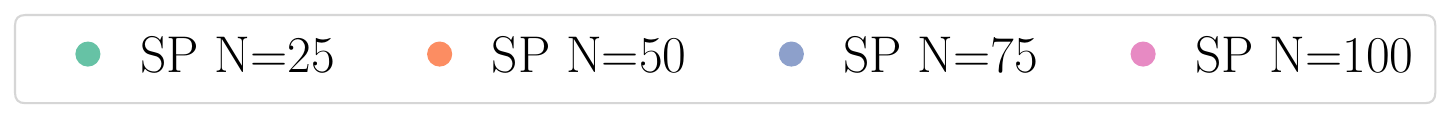}%
    \end{subfigure}
    \begin{subfigure}{\textwidth}
        \begin{tabularx}{\textwidth}{cc|c}
            \includegraphics[width=.33\columnwidth]{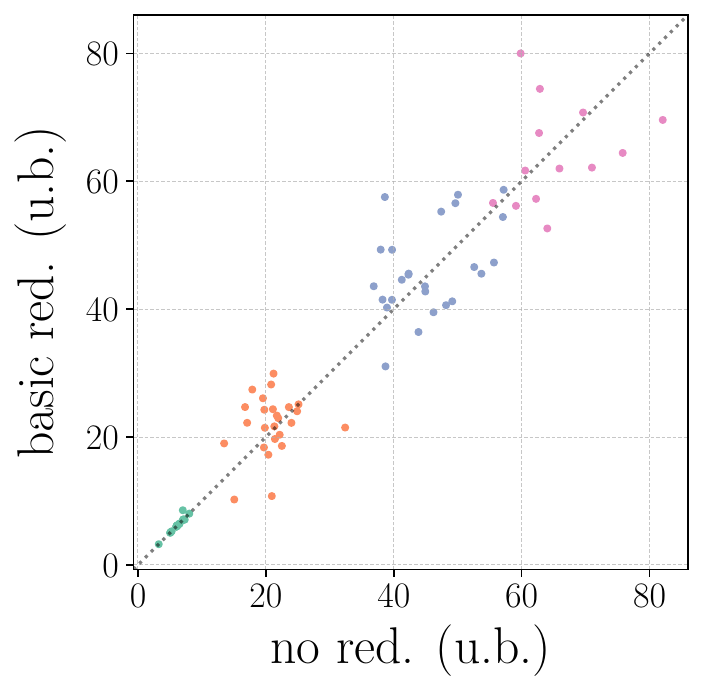}                                 &
            \includegraphics[width=.33\columnwidth]{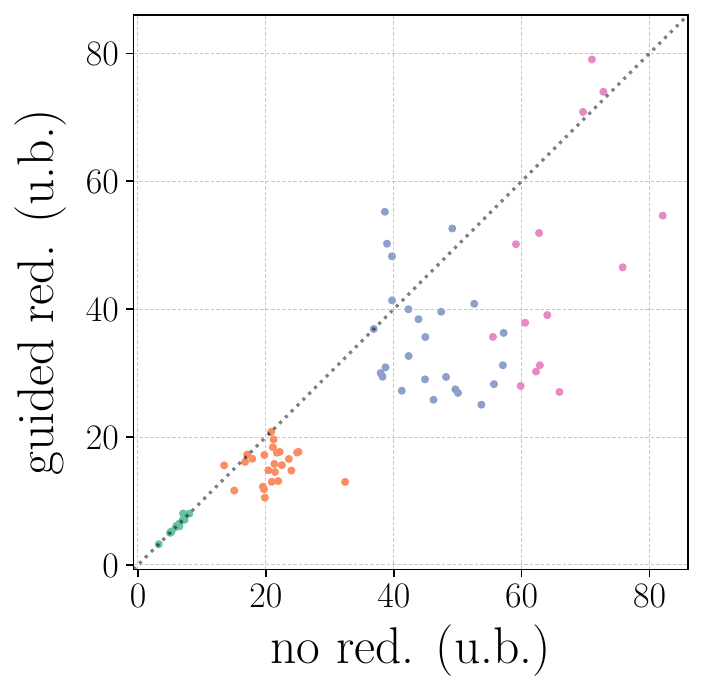}                       &
            \includegraphics[width=.33\columnwidth]{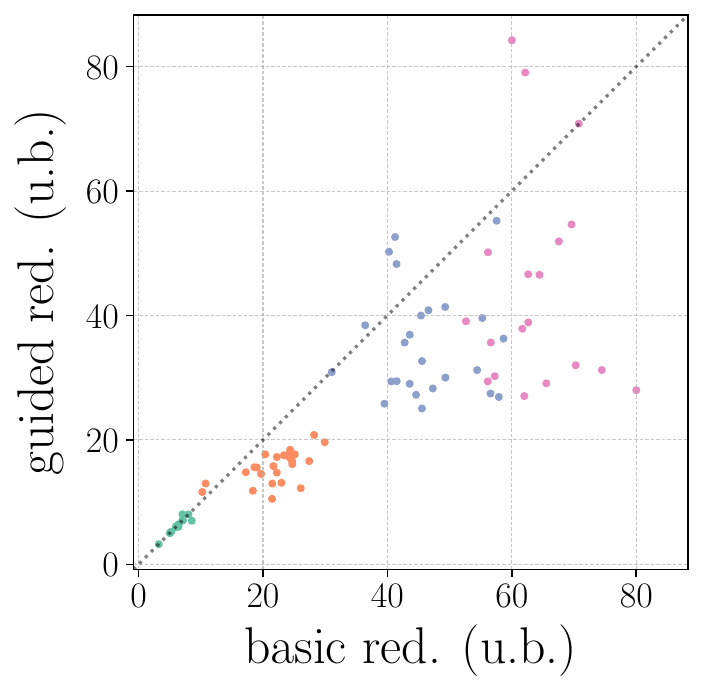}                       \\
            \includegraphics[width=.33\columnwidth]{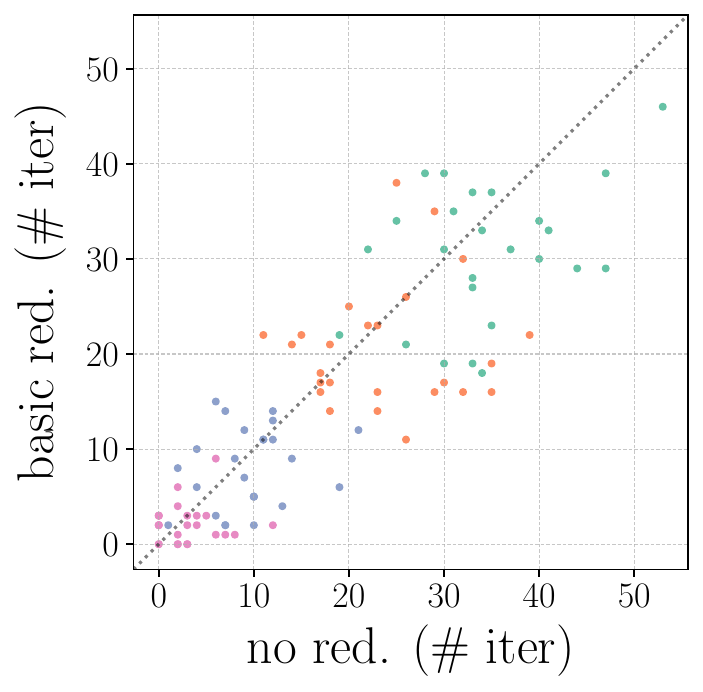}           &
            \includegraphics[width=.33\columnwidth]{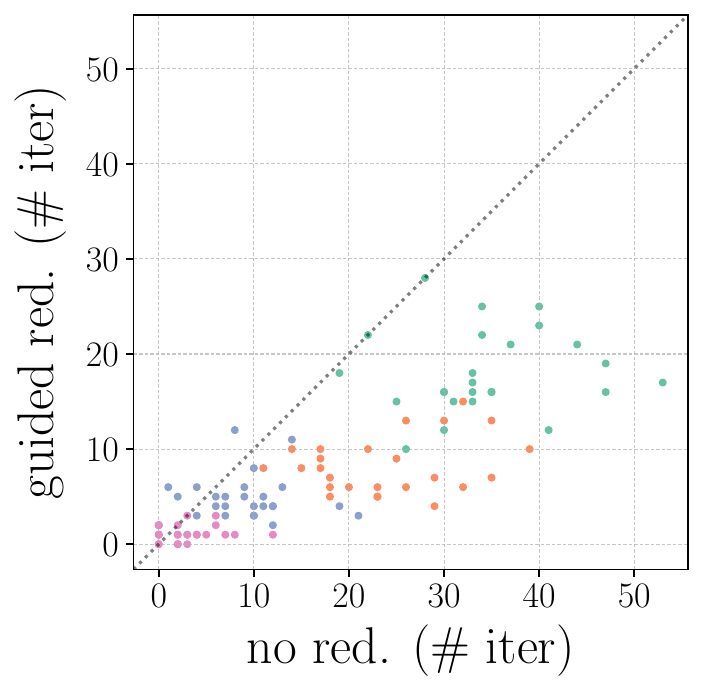} &
            \includegraphics[width=.33\columnwidth]{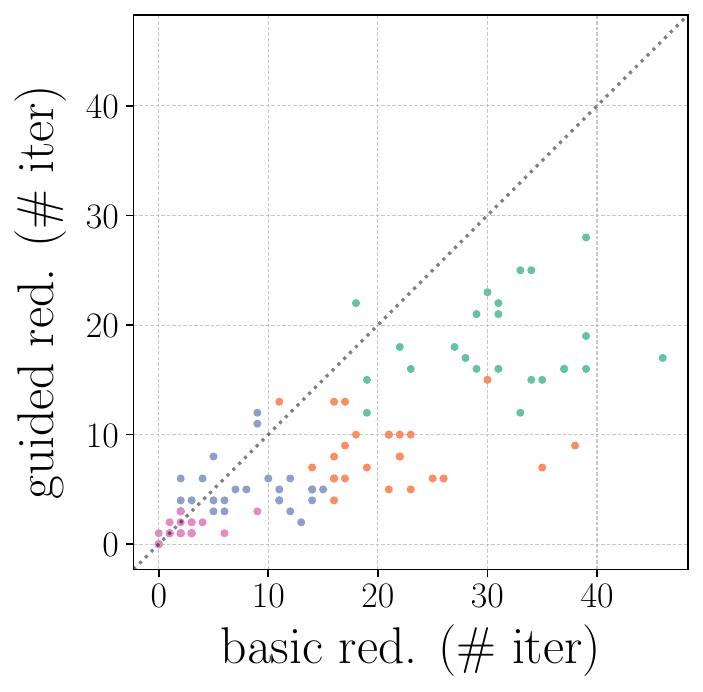}
        \end{tabularx}
    \end{subfigure}
    \caption{Results on \omlaratint-encoded strip-packing problems.}%
    \label{fig:plot:sp:lira}%
\end{figure}
\begin{figure}[t]
    \begin{subfigure}{\textwidth}%
        \centering
        \includegraphics[width=.6\columnwidth]{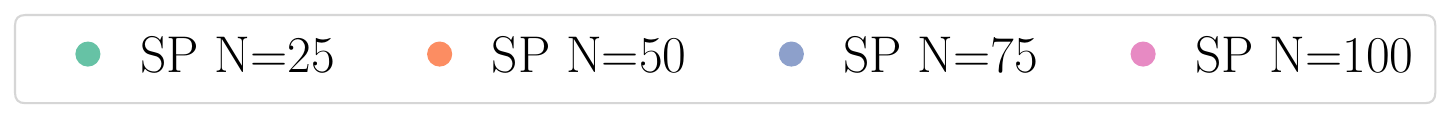}%
    \end{subfigure}
    \begin{subfigure}{\textwidth}
        \begin{tabularx}{\textwidth}{cc|c}
            \includegraphics[width=.33\columnwidth]{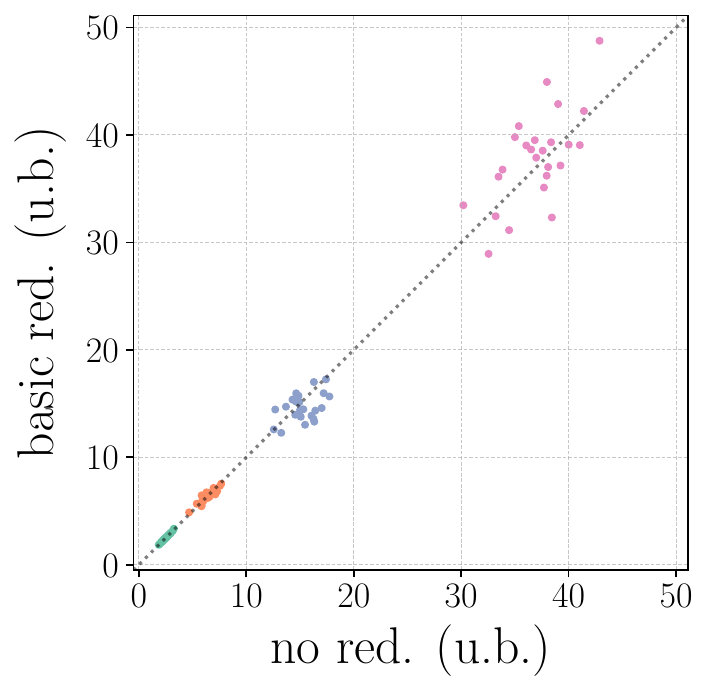}                                 &
            \includegraphics[width=.33\columnwidth]{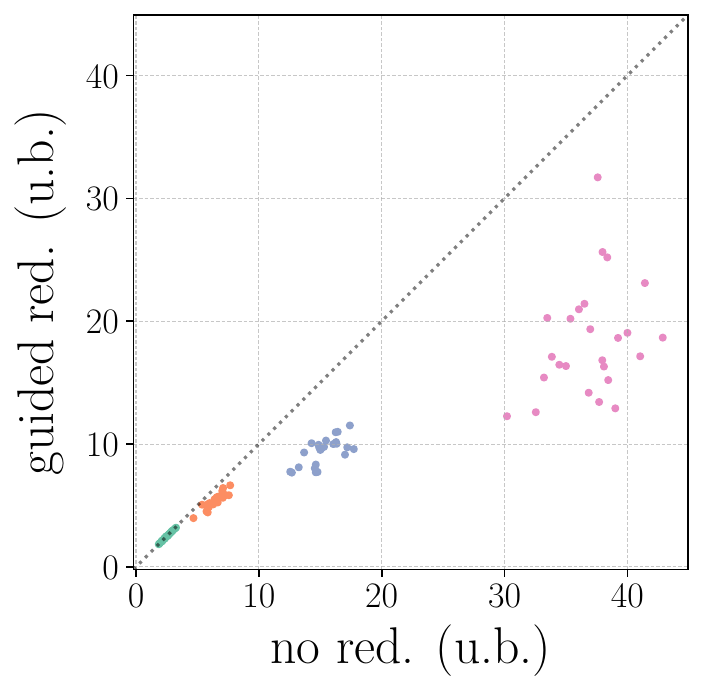}                       &
            \includegraphics[width=.33\columnwidth]{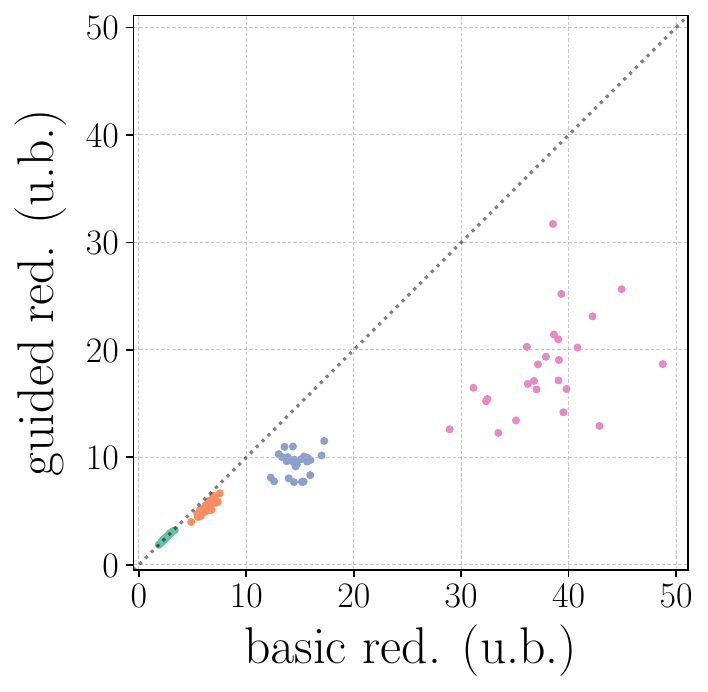}            \\
            \includegraphics[width=.33\columnwidth]{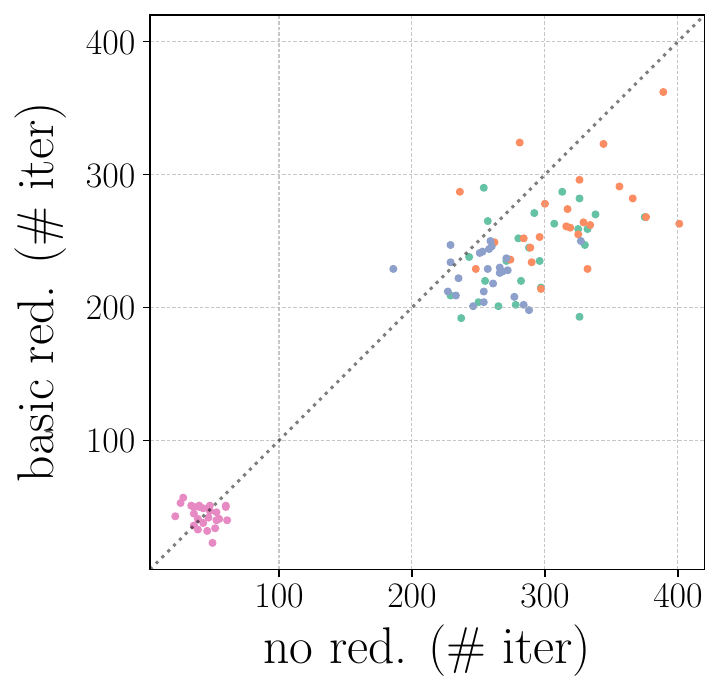}           &
            \includegraphics[width=.33\columnwidth]{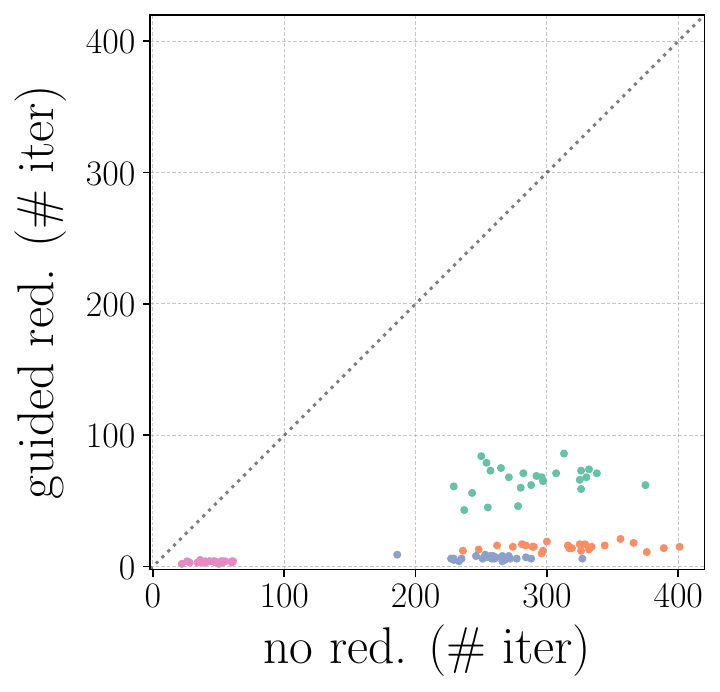} &
            \includegraphics[width=.33\columnwidth]{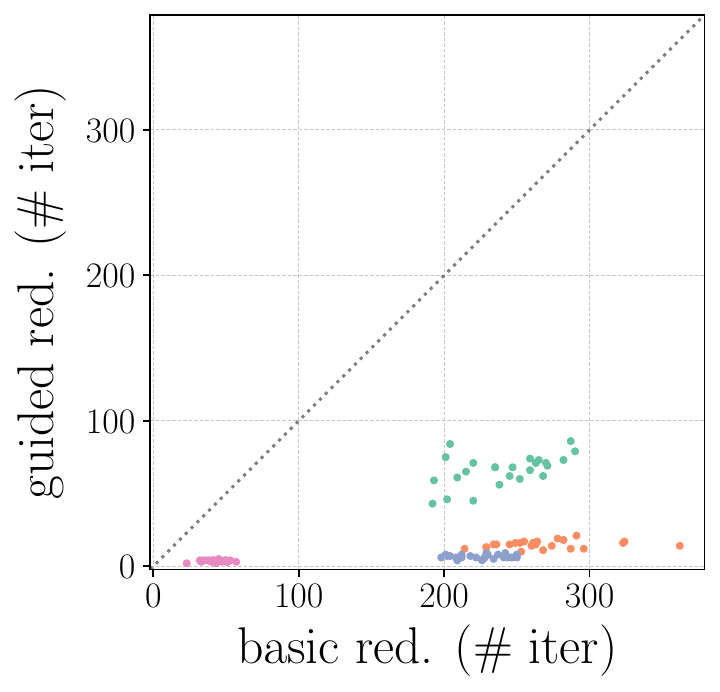}
        \end{tabularx}
    \end{subfigure}
    \caption{Results on \omlaratplusa-encoded strip-packing problems.}%
    \label{fig:plot:sp:alra}%
\end{figure}

\Cref{fig:plot:planning,fig:plot:sp:lra,fig:plot:sp:lira,fig:plot:sp:alra} show the results on temporal planning and SP benchmarks for the different theories, respectively. For each
benchmark set, we report a set of scatter plots.

On the rows, we have different metrics, namely the solving time in seconds
(time(s)), the upper bound (u.b.) ---i.e., the optimum value when the solver
terminated within the time limit, or the value of the best solution found
within the timeout otherwise--- and the number of iterations (\# iter) taken to
reach the upper bound (see \Cref{alg:omt-partial}).

On the columns, we compare the results obtained with the different
truth-assignment-reduction strategies: in the left and center columns, we
respectively compare the basic and the guided reductions with the plain
algorithm without reductions. In the right column, we compare the two reduction
strategies.

\subsubsection{\omlarat{} benchmarks.}The results are summarized in
\Cref{fig:plot:planning,fig:plot:sp:lra}.%
\paragraph{Optimal Temporal Planning (\Cref{fig:plot:planning}).}

In these benchmarks, with no truth-assignment reduction, \optimathsat{}
reported 246 timeouts out of 1520 problems, 211 with the basic reduction, and
212 with the guided reduction.

From the plots (first row, left and center columns), we can see that applying
either reduction almost uniformly improves the solving time with few
exceptions, making optimal solving up to twice as fast as with no reduction.

Moreover, we observe that reducing truth assignments is very effective also for
anytime solving (second row, left and center columns). Notice that when the
solver terminated within the timeout with both strategies, then the
corresponding points lie on the bisector, whereas when at least one strategy
times out, the points are generally below the bisector.
Indeed, this shows that, for anytime solving, both the basic and the guided
reductions allow finding a much better upper bound than with no reduction. 

Finally, we can see that both strategies are particularly effective in reducing
the number of iterations needed to either find the optimum or to reach the best
upper bound within the timeout (third row, left and center). Reducing the
number of iterations is not an advantage in itself, but it is a good indicator
of the effectiveness of truth-assignment reduction strategies in OMT.

Overall, in these benchmarks there is no clear winner between the two reduction
strategies (right column), but it is evident that applying either form of
truth-assignment reduction can be beneficial in OMT, both for optimal and
anytime solving.

\paragraph{Strip-packing (\Cref{fig:plot:sp:lra}).}
Since no instance in this set of benchmarks terminated within the timeout, for these benchmarks we omit the time plots.
We can see that here the basic reduction strategy is not really effective, since the value of the upper bound is not improved compared to the no-reduction strategy (first row, left column).
Also, the number of iterations only slightly decreases (second row, left
column), suggesting that here blindly removing atoms from the truth assignment
does not help much in finding better solutions. On the other hand, the guided
reduction strategy is much more effective, since it allows finding a much
better upper bound within the timeout (first row, center and right columns),
and the number of iterations is drastically reduced (second row, center and
right columns).

\subsubsection{\omlaratint{} benchmarks.}The results are summarized in \Cref{fig:plot:sp:lira}.%
\paragraph{Strip-packing (\Cref{fig:plot:sp:lira}).}
The plots show that also for \laratint{}, the trend is similar to the one for \larat{}.
In fact, the basic reduction strategy is not very effective for improving the upper bound (first row, left column), and the number of iterations is only slightly reduced (second row, left column).
On the other hand, the guided reduction allows finding a much better upper bound within the timeout (first row, center and right columns).
Notice that, since these problems are much harder than the \omlarat{} ones, the number of iterations completed within the timeout is much smaller, so that the upper bounds found are also bigger.
\subsubsection{\omlaratplusa{} benchmarks.}The results are summarized in \Cref{fig:plot:sp:alra}.%
\paragraph{Strip-packing (\Cref{fig:plot:sp:alra}).}
Similarly, here we can see that the technique works also for combination of theories, such as \laratplusa{}. The results are similar to the ones for \larat{}, and the advantage of the guided reduction is even more evident.

\subsubsection{Discussion.}




The results show that applying either form of truth-assignment reduction can be
beneficial in OMT, both for optimal and anytime solving, and that accurately
selecting which atoms to remove from the truth assignment can make a
significant difference in finding better solutions in fewer iterations.

We observe, however, that a much smaller number of iterations, i.e.\ of truth
assignments enumerated, not always correlates linearly with the solving time.
This can be due to several reasons.

First, we remark that global efficiency of OMT depends on several factors,
including the enumeration order of truth assignments, and different literal
selections may alter this order. Also, the removal of some literals from the
assignments can prevent the removal of others in subsequent iterations. The
effects of these factors are quite unpredictable.

Second, we notice that in these problems, the number of truth assignments
enumerated is typically contained to a few hundred. 
In fact, in OMT the bounds on the objective function already allow performing a
very effective pruning of the search space. 

Moreover, this pruning is typically done by theory reasoning, and most of it
has to be done anyway, regardless of the number of truth assignments
enumerated. Making it in a single iteration or in many iterations may not
reflect as much on the solving time, because of the efficient incrementality of
SMT solvers, which can reduce a lot the cost of consecutive iterations.
%
\section{Conclusions and Future Work}%
\label{sec:conclusions}

In this paper, we have investigated the role of truth assignment enumeration in OMT solving, and proposed some ways for exploiting partial truth assignments for improving the efficiency and effectiveness of the search. 
In particular, we have proposed a truth assignment reduction strategy that takes advantage of the properties of the optimization problem to accurately choose the atoms to remove from the truth assignment.

We have implemented the proposed strategies in the \optimathsat{} solver, and evaluated them on a set of \omlarat{}, \omlaratint{}, and \omlaratplusa benchmarks.
Our experimental results show that the proposed strategies can significantly improve the performance of the solver, uniformly reducing the overall solving time for optimal solving, and finding much better solutions for anytime solving for all the analyzed theories.  


Other truth assignment reduction strategies, such as entailment-based methods~\cite{friedEntailingGeneralizationBoosts2024}, have shown significant benefits for SAT enumeration. Their extension to OMT problems could be a promising direction for future work.

\FloatBarrier
\newpage
%
%
%
\bibliographystyle{splncs04}
\bibliography{bibliography}

\begin{thebibliography}{10}
\providecommand{\url}[1]{\texttt{#1}}
\providecommand{\urlprefix}{URL }
\providecommand{\doi}[1]{https://doi.org/#1}

\bibitem{albertGASOLGasAnalysis2020}
Albert, E., Correas, J., Gordillo, P., {Rom{\'a}n-D{\'i}ez}, G., Rubio, A.:
  {{GASOL}}: {{Gas Analysis}} and {{Optimization}} for {{Ethereum Smart
  Contracts}}. In: {{TACAS}} 2020. pp. 118--125. Springer (2020)

\bibitem{barrettSatisfiabilityModuloTheories2021}
Barrett, C., Sebastiani, R., Seshia, S.A., Tinelli, C.: Satisfiability {{Modulo
  Theories}}. In: Handbook of {{Satisfiability}}, {{FAIA}}, vol.~336, pp.
  1267--1329. IOS Press, 2 edn. (2021)

\bibitem{bianSolvingSATMaxSAT2020}
Bian, Z., Chudak, F., Macready, W., Roy, A., Sebastiani, R., Varotti, S.:
  Solving {{SAT}} (and {{MaxSAT}}) with a quantum annealer: {{Foundations}},
  encodings, and preliminary results. Inf Comput  \textbf{275},  104609 (2020)

\bibitem{bigarellaOptimizationModuloNonlinear2021}
Bigarella, F., Cimatti, A., Griggio, A., Irfan, A., Jon{\'a}{\v s}, M., Roveri,
  M., Sebastiani, R., Trentin, P.: Optimization {{Modulo Non-linear
  Arithmetic}} via {{Incremental Linearization}}. In: {{FROCOS}} 2021. pp.
  213--231. {{LNCS}}, Springer (2021)

\bibitem{bjornerNZMaximalSatisfaction2014}
Bjorner, N., Phan, A.D.: {{$\nu$Z}} - {{Maximal Satisfaction}} with {{Z3}}. In:
  Proc {{International Symposium}} on {{Symbolic Computation}} in {{Software
  Science}}. pp.~1--9 (2014)

\bibitem{bjornerNZOptimizingSMT2015}
Bj{\o}rner, N., Phan, A.D., Fleckenstein, L.: {{$\nu$Z}} - {{An Optimizing SMT
  Solver}}. In: {{TACAS}} 2015. pp. 194--199. {{LNCS}}, Springer (2015)

\bibitem{bofillEfficientSMTApproach2017}
Bofill, M., Coll, J., Suy, J., Villaret, M.: An {{Efficient SMT Approach}} to
  {{Solve MRCPSP}}/max {{Instances}} with {{Tight Constraints}} on
  {{Resources}}. In: {{CP}} 2017. pp. 71--79. {{LNCS}}, Springer (2017)

\bibitem{bozzanoEfficientTheoryCombination2006}
Bozzano, M., Bruttomesso, R., Cimatti, A., Junttila, T., Ranise, S., {van
  Rossum}, P., Sebastiani, R.: Efficient {{Theory Combination}} via {{Boolean
  Search}}. Inf Comput  \textbf{204}(10),  1493--1525 (2006)

\bibitem{mathsat5_tacas13}
Cimatti, A., Griggio, A., Schaafsma, B.J., Sebastiani, R.: The {{MathSAT5 SMT
  Solver}}. In: {{TACAS}} 2013. pp. 93--107. {{LNCS}}, Springer (2013)

\bibitem{dingEffectivePrimeFactorization2024}
Ding, J., Spallitta, G., Sebastiani, R.: Effective prime factorization via
  quantum annealing by modular locally-structured embedding. Sci Rep
  \textbf{14}(1), ~3518 (2024)

\bibitem{dutertreFastLinearArithmeticSolver2006}
Dutertre, B., {de Moura}, L.: A {{Fast Linear-Arithmetic Solver}} for
  {{DPLL}}({{T}}). In: {{CAV}} 2006. pp. 81--94. {{LNCS}}, Springer (2006)

\bibitem{friedEntailingGeneralizationBoosts2024}
Fried, D., Nadel, A., Sebastiani, R., Shalmon, Y.: Entailing {{Generalization
  Boosts Enumeration}}. In: {{SAT}} 2024. {{LIPIcs}}, vol.~305, pp.
  13:1--13:14. LZI (2024)

\bibitem{friedAllSATCombinationalCircuits2023}
Fried, D., Nadel, A., Shalmon, Y.: {{AllSAT}} for {{Combinational Circuits}}.
  In: {{SAT}} 2023. {{LIPIcs}}, vol.~271, pp. 9:1--9:18. LZI (2023)

\bibitem{henryHowComputeWorstcase2014}
Henry, J., Asavoae, M., Monniaux, D., Ma{\"i}za, C.: How to compute worst-case
  execution time by optimization modulo theory and a clever encoding of program
  semantics. In: {{LCTES}} 2014. pp. 43--52. ACM (2014)

\bibitem{lahiriSMTTechniquesFast2006}
Lahiri, S.K., Nieuwenhuis, R., Oliveras, A.: {{SMT Techniques}} for {{Fast
  Predicate Abstraction}}. In: {{CAV}} 2006. pp. 424--437. {{LNCS}}, Springer
  (2006)

\bibitem{leofanteOptimalPlanningModulo2021}
Leofante, F., Giunchiglia, E., {\'A}brah{\'a}m, E., Tacchella, A.: Optimal
  {{Planning Modulo Theories}}. In: {{IJCAI}} 2020. pp. 4128--4134 (2021)

\bibitem{liSymbolicOptimizationSMT2014}
Li, Y., Albarghouthi, A., Kincaid, Z., Gurfinkel, A., Chechik, M.: Symbolic
  optimization with {{SMT}} solvers. In: {{POPL}} 2014. pp. 607--618. ACM
  (2014)

\bibitem{maSolvingGeneralizedOptimization2012}
Ma, F., Yan, J., Zhang, J.: Solving {{Generalized Optimization Problems
  Subject}} to {{SMT Constraints}}. In: {{FAW-AAIM}} 2012. pp. 247--258.
  Springer (2012)

\bibitem{marques-silvaConflictDrivenClauseLearning2021}
{Marques-Silva}, J., Lynce, I., Malik, S.: Conflict-{{Driven Clause Learning
  SAT Solvers}}. In: Handbook of {{Satisfiability}}, {{FAIA}}, vol.~336. IOS
  Press (2021)

\bibitem{masinaCNFConversionDisjoint2023}
Masina, G., Spallitta, G., Sebastiani, R.: On {{CNF Conversion}} for {{Disjoint
  SAT Enumeration}}. In: {{SAT}} 2023. {{LIPIcs}}, vol.~271, pp. 15:1--15:16.
  LZI (2023)

\bibitem{morgadoGoodLearningImplicit2005}
Morgado, A., {Marques-Silva}, J.: Good {{Learning}} and {{Implicit Model
  Enumeration}}. In: {{ICTAI}} 2005. pp. 131--136. IEEE Computer Society (2005)

\bibitem{nadelBitVectorOptimization2016}
Nadel, A., Ryvchin, V.: Bit-{{Vector Optimization}}. In: {{TACAS}} 2016. pp.
  851--867. {{LNCS}}, Springer (2016)

\bibitem{nguyenMultiobjectiveReasoningConstrained2018}
Nguyen, C.M., Sebastiani, R., Giorgini, P., Mylopoulos, J.: Multi-objective
  reasoning with constrained goal models. Requir Eng  \textbf{23}(2),  189--225
  (2018)

\bibitem{nieuwenhuisSATModuloTheories2006}
Nieuwenhuis, R., Oliveras, A.: On {{SAT Modulo Theories}} and {{Optimization
  Problems}}. In: {{SAT}} 2006. pp. 156--169. {{LNCS}}, Springer (2006)

\bibitem{panjkovicExpressiveOptimalTemporal2023}
Panjkovic, S., Micheli, A.: Expressive {{Optimal Temporal Planning}} via
  {{Optimization Modulo Theory}}. AAAI 2023  \textbf{37}(10),  12095--12102
  (2023)

\bibitem{panjkovicAbstractActionScheduling2024}
Panjkovic, S., Micheli, A.: Abstract {{Action Scheduling}} for {{Optimal
  Temporal Planning}} via {{OMT}}. AAAI 2024  \textbf{38}(18),  20222--20229
  (2024)

\bibitem{plaistedStructurepreservingClauseForm1986}
Plaisted, D.A., Greenbaum, S.: A {{Structure-preserving Clause Form
  Translation}}. J Symb Comput  \textbf{2}(3),  293--304 (1986)

\bibitem{raviMinimalAssignmentsBounded2004}
Ravi, K., Somenzi, F.: Minimal {{Assignments}} for {{Bounded Model Checking}}.
  In: {{TACAS}} 2004. {{LNCS}}, vol.~2988, pp. 31--45. Springer (2004)

\bibitem{sebastianiLazySatisfiabilityModulo2007}
Sebastiani, R.: Lazy {{Satisfiability Modulo Theories}}. JSAT  \textbf{3}(3-4),
   141--224 (2007)

\bibitem{sebastianiOptimizationSMTLAQ2012}
Sebastiani, R., Tomasi, S.: Optimization in {{SMT}} with {{LA}}({{Q}}) {{Cost
  Functions}}. In: {{IJCAR}} 2012. {{LNCS}}, vol.~7364, pp. 484--498. Springer
  (2012)

\bibitem{sebastianiOptimizationModuloTheories2015}
Sebastiani, R., Tomasi, S.: Optimization {{Modulo Theories}} with {{Linear
  Rational Costs}}. ACM Trans. Comput. Logic  \textbf{16}(2),  12:1--12:43
  (2015)

\bibitem{sebastianiPushingEnvelopeOptimization2015}
Sebastiani, R., Trentin, P.: Pushing the {{Envelope}} of {{Optimization Modulo
  Theories}} with {{Linear-Arithmetic Cost Functions}}. In: {{TACAS}} 2015. pp.
  335--349. {{LNCS}}, Springer (2015)

\bibitem{sebastianiOptiMathSATToolOptimization2020}
Sebastiani, R., Trentin, P.: {{OptiMathSAT}}: {{A Tool}} for {{Optimization
  Modulo Theories}}. J Autom Reason  \textbf{64}(3),  423--460 (2020)

\bibitem{spallittaEnhancingSMTbasedWeighted2024}
Spallitta, G., Masina, G., Morettin, P., Passerini, A., Sebastiani, R.:
  Enhancing {{SMT-based Weighted Model Integration}} by {{Structure
  Awareness}}. Artif Intell  \textbf{328},  104067 (2024)

\bibitem{spallittaDisjointPartialEnumeration2024}
Spallitta, G., Sebastiani, R., Biere, A.: Disjoint {{Partial Enumeration}}
  without {{Blocking Clauses}}. In: {{AAAI}} 2024. vol.~38, pp. 8126--8135
  (2024)

\bibitem{spallittaDisjointProjectedEnumeration2025}
Spallitta, G., Sebastiani, R., Biere, A.: Disjoint projected enumeration for
  {{SAT}} and {{SMT}} without blocking clauses. Artif Intell  \textbf{345},
  104346 (2025)

\bibitem{tesoStructuredLearningModulo2017}
Teso, S., Sebastiani, R., Passerini, A.: Structured learning modulo theories.
  Artif Intell  \textbf{244},  166--187 (2017)

\bibitem{todaImplementingEfficientAll2016}
Toda, T., Soh, T.: Implementing {{Efficient All Solutions SAT Solvers}}. ACM J.
  Exp. Algorithmics  \textbf{21},  1--44 (2016)

\bibitem{trentinOptimizationModuloTheories2021}
Trentin, P., Sebastiani, R.: Optimization {{Modulo}} the {{Theories}} of
  {{Signed Bit-Vectors}} and {{Floating-Point Numbers}}. J Autom Reason
  \textbf{65}(7),  1071--1096 (2021)

\bibitem{tseitinComplexityDerivationPropositional1983}
Tseitin, G.S.: On the {{Complexity}} of {{Derivation}} in {{Propositional
  Calculus}}. In: Automation of {{Reasoning}}: 2: {{Classical Papers}} on
  {{Computational Logic}} 1967--1970, pp. 466--483. Symbolic {{Computation}},
  Springer (1983)

\bibitem{tsiskaridzeGeneralizedOptimizationModulo2024}
Tsiskaridze, N., Barrett, C., Tinelli, C.: Generalized {{Optimization Modulo
  Theories}}. In: Automated {{Reasoning}}. pp. 458--479. Springer (2024)

\end{thebibliography}

\end{document}